# Investigating Flight Envelope Variation Predictability of Impaired Aircraft using Least-Squares Regression Analysis

Ramin Norouzi,[1] Amirreza Kosari,[2] and Mohammad Hossein Sabour[3]
*University of Tehran, Tehran, 1439957131, Iran*

**Aircraft failures alter the aircraft dynamics and cause maneuvering flight envelope to change. Such envelope variations are nonlinear and generally unpredictable by the pilot as they are governed by the aircraft's complex dynamics. Hence, in order to prevent in-flight Loss of Control it is crucial to practically predict the impaired aircraft's flight envelope variation due to any a-priori unknown failure degree. This paper investigates the predictability of the number of trim points within the maneuvering flight envelope and its centroid using both linear and nonlinear least-squares estimation methods. To do so, various polynomial models and nonlinear models based on hyperbolic tangent function are developed and compared which incorporate the influencing factors on the envelope variations as the inputs and estimate the centroid and the number of trim points of the maneuvering flight envelope at any intended failure degree. Results indicate that both the polynomial and hyperbolic tangent function-based models are capable of predicting the impaired fight envelope variation with good precision. Furthermore, it is shown that the regression equation of the best polynomial fit enables direct assessment of the impaired aircraft's flight envelope contraction and displacement sensitivity to the specific parameters characterizing aircraft failure and flight condition.**

## Nomenclature

$V$ = total airspeed, knot

$\alpha$ = angle of attack, deg

$\beta$ = sideslip angle, deg

---

[1] Ph.D. Candidate, Faculty of New Sciences and Technologies; ramin.norouzi@ut.ac.ir. Student Member AIAA.
[2] Associate Professor, Faculty of New Sciences and Technologies; kosari_a@ut.ac.ir.
[3] Assistant Professor, Faculty of New Sciences and Technologies; sabourmh@ut.ac.ir.

| Symbol | | Description |
|---|---|---|
| $p, q, r$ | = | angular velocity components (roll rate, pitch rate, yaw rate, respectively), deg/s |
| $\phi, \theta, \psi$ | = | Euler angles (roll, pitch, yaw, respectively), deg |
| $\gamma$ | = | flight path angle, deg |
| $x$ | = | state vector |
| $u$ | = | control vector |
| $\delta_e$ | = | elevator deflection angle, deg |
| $\delta_a$ | = | aileron deflection angle, deg |
| $\delta_r$ | = | rudder deflection angle, deg |
| $\delta_{th}$ | = | throttle setting (%) $\in [0, 1]$ |
| $e$ | = | fitted model's residual vector |
| $\hat{y}$ | = | fitted model's outcome vector |
| $\eta$ | = | fitted model's parameters (coefficients) vector |
| $z$ | = | input vector with the input factors $(h, \gamma, LL, UL)$ as its components |
| $h$ | = | flight altitude, ft |
| $UL$ | = | upper limit of the rudder deflection angle, deg |
| $LL$ | = | lower limit of the rudder deflection angle, deg |
| $n_{trim}$ | = | number of the trim points within the maneuvering flight envelope |
| $C_{trim}$ | = | centroid of the maneuvering flight envelope |
| $d$ | = | total degree of polynomial |
| Poly2222 | = | 2$^{nd}$ degree polynomial (all input variables have a maximum degree of 2) |
| Poly3333 | = | 3$^{rd}$ degree polynomial (all input variables have a maximum degree of 3) |
| Poly4444 | = | 4$^{th}$ degree polynomial (all input variables have a maximum degree of 4) |
| Poly3344 | = | 4$^{th}$ degree polynomial (maximum degrees of $h, \gamma$ are 3, maximum degrees of $LL, UL$ are 4) |
| $\mathbb{g}$ | = | gradient of the objective function $F$ |
| $\mathbb{H}$ | = | Hessian of the objective function $F$ |
| $\mathbb{J}$ | = | Jacobian of the objective function $F$ |
| $\mathbb{I}$ | = | identity matrix |
| $\mathbb{d}$ | = | Newton's method search direction |
| $\xi$ | = | Marquardt parameter |
| $\mathcal{T}$ | = | hyperbolic tangent function |
| $S^1$ | = | total number of neurons in the first (hidden) layer |
| $S^2$ | = | total number of neurons in the second (last) layer |
| $\mathbb{K}$ | = | backpropagation training iteration |
| $a_\mathcal{L}^\mathbb{C}$ | = | net output of $\mathcal{L}^{th}$ neuron in layer $\mathbb{C}$ |
| $n_\mathcal{L}^\mathbb{C}$ | = | net input of $\mathcal{L}^{th}$ neuron in layer $\mathbb{C}$ |
| $b_\mathcal{L}^\mathbb{C}$ | = | bias of $\mathcal{L}^{th}$ neuron in layer $\mathbb{C}$ |
| $W_{\mathcal{L},\mathcal{J}}^\mathbb{C}$ | = | weight's element of $\mathcal{L}^{th}$ neuron in layer $\mathbb{C}$ due to $\mathcal{J}^{th}$ neuron of the previous layer |
| $\lambda^\mathbb{C}$ | = | standard backpropagation sensitivity |
| $\hat{\lambda}^\mathbb{C}$ | = | Marquardt sensitivity |
| $e_{\mathcal{E},i}$ | = | contribution of the $\mathcal{E}^{th}$ neuron of the last (second) layer to the error of the $i^{th}$ training sample |
| $E_{z_j}$ | = | statistical expected value of the $j^{th}$ input factor |
| $\mathbb{S}_j$ | = | first-order sensitivity index, main effect of the $j^{th}$ input factor on the model output |
| $\mathbb{S}_{ji}$ | = | higher order sensitivity index, joint effect of the $j^{th}$ and $i^{th}$ input factors on the model output |
| $\mathbb{S}_j^T$ | = | total-order sensitivity index, total effect of the $j^{th}$ input factor on the model output |

Subscripts

| | | |
|---|---|---|
| $m, \mathcal{P}$ | = | total number of training samples, total number of model parameters, respectively |

## I. Introduction

According to the statistical report published by Boeing in October 2018, in-flight Loss of Control (LOC); with 14 accidents and a total of 1129 fatalities, is the primary contributor among different factors leading to the fatal accident of commercial airliners over the years 2008-2017 [1]. Another report published by UK Civil Aviation Authority in 2013 investigating the fatal accidents of 2002-2013 shows that almost 40% of all fatal accidents were related to the loss of control, making it the major cause of the accidents [2]. A noteworthy observation in this period is that there has been a decreasing trend in the number of fatal accidents despite the increase in the number of flights, prominently due to the emerging of more accurate flight control and safety systems and more intelligent control automation systems [2]. However, still LOC holds the greatest share in fatal accidents, despite all improvements made to pilot training and aircraft systems. LOC usually occurs following an upset condition which can be caused by technical failures such as control surface defects, or external events such as icing, or internal sources such as pilot inputs, or a combination of these factors [3].

In case of technical failures or external events, aircraft dynamics and parameters are changed, and so are the flight envelope and its kinematic constraints. Generally, the degraded performance of an impaired aircraft which is dictated by its altered nonlinear dynamics is characterized by the new flight envelope which is more confined than the nominal flight envelope of the unimpaired aircraft [4]. Aircraft new dynamics and new flight envelope boundaries are not determined for the pilot, because pilots are not and cannot be trained for all possible failure situations. So the pilot, who is under stress and excess workload, tries to plan a safe landing trajectory as soon as possible, which may include an input or maneuver outside the new admissible flight envelope, leading to LOC [4]. So far, most systems designed in response to aircraft failures are those incorporating different adaptive or reconfigurable control methods. Whilst these controllers are necessary to stabilize the impaired aircraft and maintain its controllability, they cannot guarantee the pilot or the autopilot that the sequence of maneuvers (states) they have chosen is feasible based on the new altered dynamics of the impaired aircraft [5]. Hence, in order to handle an aircraft with degraded performance, in addition to an adaptive/reconfigurable controller, a system characterizing the reduced performance of the aircraft is required to augment the Flight Management System (FMS) [6].

Therefore, the main challenge in the prevention of LOC-led-accidents is to increase the pilot's situational awareness and develop better FMS augmentation systems, which both require post-failure flight envelope characterization. This is challenging as the aircraft damages impose additional nonlinear influences on the stability and control from dynamic motions [7].

Various methods have been used in previous researches to evaluate the flight envelope. A flight envelope is defined as a set of attainable trim states within a set of constraints. Loss of control may occur, once any of the constraints is violated [8]. Being attainable means the aimed trim state belongs to the region of attraction (ROA) of the current equilibrium state and the new stability margin is adequate for impaired aircraft to cope with disturbances such as gust [5]. Hence, one way to evaluate the aircraft flight envelope is to determine the ROA of equilibrium points in the nonlinear system. There are different methods proposed for evaluating the ROA. For instance, in [9], a Lyapunov function method has been used to estimate the attraction region of a stable equilibrium point in a nonlinear system, whereas in [10], the proposed method is based on topological properties of autonomous nonlinear dynamical systems. Linear Matrix Inequality (LMI) theory which was introduced in [11] to deal with nonconvex problems has been used in a number of other researches to find ROA, as in [12], where subsets of the ROA are computed using real algebraic geometry theory via reformulating the problem as an LMI. In [13], a method to evaluate the attraction region of equilibrium points of quadratic systems was proposed in which a certain box was assessed to see whether it belongs to the ROA. Also in [14], linear reachable set and nonlinear ROA techniques were used to develop algorithms to assess the dynamic flight envelope of the NASA Generic Transport Model (GTM). Recently, stable manifold theory was employed to construct the ROA representing dynamic flight envelope [15, 16].

Another group is comprised of methods estimating the safe flight envelope (which is the intersection of forward and backward reachable sets) by solving a reachability problem. Methods within this group calculate the reachability-based safe envelope by transforming the problem into Hamilton-Jacobi partial differential equations and solving it with level set methods [17, 18, 19], including a semi-Lagrangian method [20].

Other methods evaluate the trim envelope (which is a subset of the safe flight envelope) by directly computing the achievable trim points based on high-fidelity models. For instance, the interval analysis method is used to derive trim states in [21], whereas in [22], steady states are calculated using Newton-Raphson method to evaluate the steady performance and maneuvering capabilities of an unimpaired aircraft in helical trajectories. Specifically, the method to evaluate 3D maneuvering flight envelopes of an impaired aircraft based on calculating all trim points in a point-by-point schema was first introduced by M. J. Strube and E. M. Atkins in 2004 in [23], and elaborated in 2005 in [24]. Since then, it has been the basis and part of several studies associated with flight envelope estimation of impaired aircraft and post-failure path planning, such as [5, 6, 25-28]. The method defines trim points as steady state maneuvers characterized by velocity, climb rate, turn rate and altitude. It derives trim points by simultaneously minimizing the 6 degree-of-freedom (DOF) nonlinear equations of motion based on new dynamics of the damaged aircraft. In [6, 26, and 27], this method was applied to the NASA GTM with left wing damage to evaluate post-damage flight envelopes. Trim points derived via this method can then be selected as motion primitives to form a feasible trajectory in post-failure motion planning. This idea was first introduced in 2002 in [29], and was followed in many other studies such as [5, 6, 25, 27, and 28]. Alternative methods to evaluate flight envelope include utilizing bifurcation analysis [30], and continuation technique [31] where instead of trimming every steady state point, the boundary of the impaired aircraft's flight envelope is directly computed using a continuation technique.

In order to prevent LOC, it is important to evaluate the envelope variation by calculating the key characteristics of the impaired aircraft's new flight envelope in real-time after the occurrence of the failure. However, estimating the entire flight envelope of the impaired aircraft based on a high-fidelity nonlinear model using the aforementioned methods and specifying the characteristics of the evaluated envelope is computationally intensive and practically impossible to be implemented onboard the aircraft for an unpredicted failure. For instance, the "curse of dimensionality" [32] associated with the reachability set method limits its online application to low-dimensional problems [33]. Therefore, researchers have adopted various approaches to tackle this challenge. One approach is to create an offline database of flight envelopes for a priori known failures and deploy it onboard the aircraft. For instance, an augmentation to the conventional FMS is proposed in [34] with the aim of preventing loss of control which assumes the offline database is applicable to any specific case online. A similar concept is considered in [19] with application to actuator faults, and in [33, 35-37] with extension to structural damages where a database of flight envelopes corresponding to the most often occurring failures is created offline and accessed onboard the aircraft to retrieve the closest envelopes to the occurred failure degree. The retrieved envelopes are then interpolated to estimate the actual impaired envelope. The results of the offline database approach are only accurate if enough envelopes have been evaluated to be stored in the database. Hence, the drawback with this approach is that it requires carrying massive databases onboard. This is specifically important as the graphical results presented in [37] depict the interpolation output of two candidate envelopes which seem to be almost isotropic scale of each other. However, this is not always the case considering the nonlinear dynamics of the aircraft which directly influence the envelope variations at different failure degrees. Therefore in this approach, numerous envelopes with appropriate distribution on ranges of the failure degree and other affecting factors should be evaluated and added to the database.

Utilizing reduced complexity models instead of high-fidelity models is another approach which enables efficient and fast flight envelope estimation onboard the aircraft. Time scale separation is used in [38, and 39] to develop a computationally efficient model based on the variables that drive the slow dynamics of the aircraft state, whereas a point mass dynamic model along with a differential vortex lattice algorithm is used in [40] to estimate the impaired aircraft flight envelopes. While their online implementation is feasible, flight envelopes estimated based on reduced complexity models are considerably simplified [34] and lack specific steady state maneuvers' characteristic.

A third approach is to estimate local flight envelopes instead of the entire flight envelope. This approach has been employed in studies associated with developing adaptive flight planners where local envelopes are estimated online progressively as new flight conditions are visited [5, 28]. Linear discrete-time models are used in [41] whereas [19] confines the optimization space to speed up the reachable set calculations. A shortcoming with this approach is that the local envelope characteristics do not generally reflect the entire flight envelope characteristics of the damaged aircraft. In other words, a local envelope might be the intersection of the unimpaired and impaired flight envelopes hence being unchanged even after the occurrence of the failure and implying to the pilot that the failure has not affected the set of achievable trim points, whilst the other parts and boundaries of the global envelope might have been shrunk and even drifted within the steady-state-space.

While the aforementioned methods have their own pros and cons, an alternative approach could be developing a comprehensive mathematical model relating the main parameters causing envelope variation to the principal characteristics of the flight envelope which determine envelope alteration. Such a model if built based on a database of high-fidelity flight envelopes resembles the nonlinear and complex dynamics of the impaired aircraft and eliminates the need to carry a massive database onboard. Since the model is developed offline it can be fine-tuned by changing the model specifications or adding more failure cases to the training database until an acceptable level of accuracy is achieved. Also, it provides the envelope characteristics at the intended failure degree instantaneously, eliminating the time required to explore an onboard database, to retrieve and interpolate flight envelopes, and to specify the required characteristics. This is important because immediate knowledge of the altered flight envelope characteristics of the impaired aircraft is crucial to prevent loss of control and without such computationally efficient model it is impossible to acquire such knowledge for the unpredicted (i.e. a-priori unknown) failures in real-time. More importantly, the developed model which provides a direct analytical equation can be used as an efficient and precise emulator in numerous model evaluations required for analyzing the sensitivity of the envelope variation to the designated input parameters.

This paper investigates the feasibility of this approach. It should be noted that evaluating the flight envelopes of impaired aircraft with structural damages demands carrying out wind tunnel tests or implementing specific numerical methods which are beyond the scope and available resources of this research. Hence, a database previously created by the authors comprising of high-fidelity flight envelopes of an impaired GTM with actuator failures was used in this study to develop the intended nonlinear models by linear and nonlinear least-squares techniques. Also, this study is not focused on the fault detection process and it is assumed that an appropriate method such as [42] can be used to detect and isolate the control surface fault.

The number of achievable trim points within the envelope's boundary changes with different failure degrees as each failure shrinks the envelope to a different extent. Also, the remaining flight envelope and subsequently its centroid relocate with different failure degrees. These two key parameters (i.e. the impaired aircraft's flight envelope's centroid and number of trim states) which characterize the envelope variation of an impaired aircraft are of different natures and neither of the two can be estimated based on the knowledge of the other parameter. In other words, the amount of envelope contraction due to failure cannot be evaluated from the amount of envelope relocation due to the same failure and vice versa. Hence, these two parameters are selected as the desired outputs of the developing models as they provide good indications of the developed models' capability in predicting both of the two different aspects of envelope variation due to failure.

The rest of this paper is arranged into sections as follows: Sec. II introduces the dynamic model used in this research. Sec. III presents the evaluation procedure and specifications of the flight envelope database employed in this study. In Sec. IV, first, polynomial models with different degrees are developed by QR decomposition using a training dataset and their generalization ability is assessed based on their fit to a test dataset. Then, for the purpose of comparison, the same training data is used to construct nonlinear hyperbolic tangent function-based models of different sizes by solving nonlinear least-squares optimization problems with the Trust-region-reflective and Levenberg-Marquardt algorithms, and it is shown why solving the considered nonlinear least-squares regression problem can be implemented more effectively and efficiently within the neural network training framework. Hence, multiple neural networks are trained using the Levenberg-Marquardt algorithm and those with the best performance on the unseen data are selected, and results from the developed polynomials and selected neural networks are compared in detail. Finally, in Sec. V, a variance-based global sensitivity analysis is performed using the estimated polynomial regression equation to evaluate the levels of influence of the input factors on the impaired flight envelope variation. Lastly, Sec. VI concludes the paper.

## II. Dynamic Model

A key element in every scientific research is a valid model. The Generic Transport Model (GTM) – with tail number T2, is a 5.5% twin – turbine powered, dynamically scaled aircraft which is designed with the aim of flying into drastic upset conditions and being safely recovered. Extensive wind tunnel tests were performed on GTM to create an extended-envelope aerodynamic data set. Test data were obtained at angles of attack as low as $-5°$ and up to $+85°$ and sideslip angles ranging from $-45°$ to $+45°$ [43]. The GTM-T2 properties are shown in Table 1.

**Table 1 GTM-T2 Properties**

| Property | Quantity |
| --- | --- |
| Takeoff weight, $W_0$ | 257 N (26.2 kg) |
| Wing area, S | 0.5483 m$^2$ |
| Wing span, d | 2.09 m |
| Length, l | 2.59 m |
| Mean aerodynamic chord, $\bar{c}$ | 0.2790 m |

The dynamic model used in this research is the GTM-T2, high fidelity, nonlinear, 6 DOF, MATLAB – Simulink model, also known as *"GTM-DesignSim"* [44]. The model utilizes extensive wind tunnel test data in tabular form as the required aerodynamic database. Further details on the GTM can be found in [43, 45] and their references.

## III. Database of Damaged GTM's Flight Envelopes

As mentioned earlier, a previously generated database [46] including the flight envelopes of the unimpaired GTM and the impaired GTM with rudder failures is used in this research. Envelopes within this database are trim envelopes as per definition and explanations provided in the previous section. In related studies, this form of trim envelopes is also called maneuvering flight envelopes (MFEs) as they are boundaries containing steady state maneuvers (i.e. trim points). The same terminology is used in this research. It should be noted not to confuse these maneuvering flight envelopes with the reachability-based safe flight envelopes as in some researches they are referred to as maneuvering flight envelopes too. The following subsection briefly describes the computational procedure by which the MFEs of the database have been evaluated. For more details refer to [5, 24].

### A. Maneuvering Flight Envelope

A steady state maneuver is considered as the condition in which each individual force and moment is zero or constant whilst the net acting force and moment are zero [47]. This requires all linear and angular velocity rates and aerodynamic angles rates to be zero, whilst controls are fixed. Therefore, in the wind-axes coordinate system:

$$(\dot{p}, \dot{q}, \dot{r}) = (\dot{V}, \dot{\alpha}, \dot{\beta}) = 0 \tag{1}$$

(1), is a general definition of trim state. Hence, additional constraints need to be imposed to define the intended level/climbing/descending rectilinear and level/climbing/descending turning maneuvers:

$$\text{Level rectilinear: } \dot{\phi}, \dot{\theta}, \dot{\psi}, \gamma = 0 \tag{2}$$

$$\text{Climbing/descending rectilinear: } \dot{\phi}, \dot{\theta}, \dot{\psi} = 0, \ \gamma = cte \tag{3}$$

$$\text{Level turn: } \dot{\phi}, \dot{\theta}, \gamma = 0, \ \dot{\psi} = cte \tag{4}$$

$$\text{Climbing/descending turn: } \dot{\phi}, \dot{\theta} = 0, \ \dot{\psi}, \gamma = cte \tag{5}$$

Where $cte$ means a constant value which is determined based on the specific steady state maneuver characteristics (i.e. climb rate and turn rate). Rewriting the trim state definition in terms of equations of motion:

$$\dot{x}_{trim} = f(x_{trim}, u_{trim}) = 0 \tag{6}$$

$$x_{trim} = [V, \alpha, \beta, p, q, r, \phi, \theta]^T \tag{7}$$

$$u_{trim} = [\delta_{th}, \delta_e, \delta_a, \delta_r]^T \tag{8}$$

$$(\gamma = \gamma^*), \ (\dot{\psi} = \dot{\psi}^*), \ (\dot{\phi}, \dot{\theta} = 0) \tag{9}$$

In (9), $\gamma^*, \dot{\psi}^*$ are the desired constant values which define the steady state maneuvers shown in (2)–(5). Maneuvering flight envelopes are comprised of trim states characterized by four parameters $(h^*, V^*, \gamma^*, \dot{\psi}^*)$. Hence, these flight envelopes can be depicted as three-dimensional volumes at each constant flight altitude $h^*$. Each trim condition must satisfy the aircraft equations of motion, as in (6). Therefore, for each desired steady state maneuver, trim vectors $(x_{trim}, u_{trim})$ are found by simultaneously solving all aircraft nonlinear equations of motion ($\dot{x}_{trim} = 0$) for the desired flight path angle and turn rate $(\gamma^*, \dot{\psi}^*)$ at a specific total airspeed $V^*$ and a specific altitude $h^*$.

Solving ($\dot{x}_{trim} = 0$) is not analytically possible, hence the feasible trim points are derived by numerically solving the corresponding constrained nonlinear optimization problem in which the cost function is [6, 24, and 47]:

$$J(x, u) = 1/2 \, \dot{x}_{trim}^T \mathcal{G} \dot{x}_{trim} \tag{10}$$

where $\mathcal{G}$ specifies the state derivatives' contributions to the cost function $J$ which is subject to the following equality constraints (11–13), to the inequality constraints which are dictated by the physical limits on the control inputs (14), and to the inequality constraints on the bank angle, angle of attack and the flight path angle (15–17).

(11) constraints the flying altitude, total airspeed, flight path angle, and turn rate to the desired trim state values. (12) is obtained once the rate-of-climb (i.e. $V \sin \gamma$) constraint is solved for $\theta$ [47]. For a wings-level, non-sideslipping flight (12) reduces to $\theta = \alpha + \gamma$. The three equations of (13) are derived from the rotational kinematics once the steady-state maneuvers' constraint of $\dot{\phi} = \dot{\theta} = 0$ is applied. In (14), control surfaces' deflection limits are all in degree and have been set based on the values in the *GTM-DesignSim*. Also throttle is in percent (i.e. [0, 1]).

$$h - h^* = 0 \ , \ V - V^* = 0 \ , \ \gamma - \gamma^* = 0 \ , \ \dot{\psi} - \dot{\psi}^* = 0 \tag{11}$$

$$tan\theta - \frac{ab + sin\gamma^*\sqrt{a^2 - sin^2\gamma^* + b^2}}{a^2 - sin^2\gamma^*} = 0, \ \theta \neq \pm\pi/2 \tag{12}$$

where, $\quad a = cos\alpha cos\beta, \ b = sin\phi sin\beta + cos\phi sin\alpha cos\beta$

$$p + \dot{\psi}^* sin\theta = 0$$
$$q - \dot{\psi}^* cos\theta sin\phi = 0 \tag{13}$$
$$r - \dot{\psi}^* cos\theta cos\phi = 0$$

$$|\delta_{th} - 0.5| \leq 0.5 \quad |\delta_e| \leq 30$$
$$|\delta_a| \leq 20 \quad |\delta_r| \leq 30 \tag{14}$$

$$-30° \leq \phi \leq 30° \tag{15}$$

$$\alpha \leq 10.5° \tag{16}$$

$$-5° \leq \gamma \leq 5° \tag{17}$$

Being feasible is not enough for a trim state to include it inside the boundaries of the MFE. Feasibility is the necessary condition whilst stability is the sufficient condition for inclusion in the flight envelope. Stable trim points are more preferable because the aircraft naturally tends to damp the effect of small disturbances around them, while at unstable trim points; aircraft diverges away from the trim state. However, if the feasible trim state is unstable, it is accepted as part of the flight envelope if it is controllable, i.e. if the linear perturbation system about this trim state has a full rank controllability matrix. A nonlinear system is considered stable at a specific trim point, if the system inherently converges to the trim state when being in the vicinity of the trim point [48]. Hence, to evaluate the stability of the aircraft at each specific trim point $x^*$, aircraft motion is approximated by linearizing the equations of motion about $x^*$ via a linear perturbation method. A trim state is considered stable if the state matrix has no positive real eigenvalues or complex eigenvalues with positive real part. If the investigated trim point $x^*$ is unstable the controllability matrix of the perturbation system about $x^*$ is formed to check its rank. In case of a full rank matrix, $x^*$ is controllable and is included in the MFE, otherwise it will be excluded.

For commercial transport aircraft, it is more preferable to use bank angles of less than 30 degrees in path planning, especially in final approach and landing [6]. This enables shallow turns which impose small $g$-forces (up to $1.2g$) on passengers. It is also common in literature to impose such a constraint if the study model is a conventional civil airliner. For instance, in [26], maneuvering flight envelopes of a wing damaged GTM were estimated for bank angles constrained to ±20°, or in [38], a 35° bank constraint was imposed on the nominal and off-nominal flight envelope determination of a jet transport model. Therefore, a 30° bank constraint was imposed on the trim points derivation of the database used in this research. It also should be noted that the stall constraint on the bank angle is adhered to in the MFEs utilized in this study. This is evident in figures 1(a) and 2 by the gradual increase in the total airspeed value of the trim points at the lower boundary of the unimpaired MFE (i.e. stall speed) with the increase of the turn rate (bank angle). Moreover, the structural loading constraint on the bank angle is taken care of by the 30° bank limit considered in this research.

Also, an additional angle of attack constraint was imposed to prevent the optimization algorithm from trimming the aircraft at the stall or post-stall flight conditions. As mentioned earlier, GTM aerodynamic dataset includes aerodynamic data for high values of angle of attack up to +85°. This is because GTM was designed to investigate stall and post-stall regimes in the extended flight envelope regions. However, based on the results of the bifurcation analysis of GTM in [3], from $\alpha = 10.5°$, the aircraft enters an undesirable regime with steep helical spirals which are considered as upset conditions and must be recovered from, by stall recovery procedures. Hence, in the utilized database, the upper limit of $\alpha$ was constrained to 10.5°.

Another important factor to be considered is the maximum speed constraint due to structural loads and its relation with the lower limit of the flight path angle. In this study, aircraft's maximum speed ($V_{max}$) is referred to as the intersection of the thrust-available and thrust-required curves in the thrust-speed plane. These curves have two intersections which the intersection with greater speed corresponds to the maximum speed of the aircraft. At each specific altitude, the thrust-available curve is fixed but the thrust-required curve varies with different flight path angles. Increasing the negative flight path angle (i.e. steeper descent) enlarges the component of the weight along the velocity vector's axis, reduces the amount of the required thrust (i.e. lowering the thrust-required curve), and consequently results in higher maximum speed. However, such higher speed should not exceed the aircraft's dive speed ($V_D$) because at speeds higher than the dive speed, the dynamic pressure is higher than the design value of the aircraft and destructive phenomena such as wing divergence, flutter, and aileron reversal may occur and lead to structural failure or disintegration [49]. Therefore, the lower limit of the flight path angle must be chosen such that the resulting maximum speed is not bigger than the $V_D$. This requires knowledge of the exact value of the $V_D$, however regulations only provide relations for the minimum possible value of the $V_D$ and its exact value must be provided by the designer, which in the case of the GTM such information is not provided in the *GTM-DesignSim* or the related documents. Therefore, in this study, the lower limit of the flight path angle is considered to be $(-5°)$ and it is assumed that for the level flight $(\gamma = 0°)$ and for the descending flight up to $(-5°)$, $V_{max}$ is lower than the $V_D$. This assumption seems reasonable as $(-5°)$ flight path angle is in the range of typical descent angles of the commercial airliners.

It should be noted that the imposed constraints also limited the number of potential trim points that needed to be investigated by the explained numerical optimization, thus made the database generation process computationally affordable. However, the regression and sensitivity analysis of this research can be applied to other databases generated observing other constraints as well.

**B. MFE Database Specifications**

At each fixed altitude $h^*$, maneuvering flight envelopes are 3D volumes $(V, \gamma, \dot{\psi})$. To evaluate 3D MFEs of the database, the computational procedure presented in the previous subsection has been followed accordingly for each triplet $(V^*, \gamma^*, \dot{\psi}^*)$.

There are four categories of control surface failures [50]:

- Surface jam
- Control restriction, in which, upper and/or lower limits of deflection are changed to new, equal or non-equal values
- Reduced rate limits, in which, upper and/or lower rate limits are changed
- Surface runaway, which at first shows up as reduced rate limits but eventually changes to surface jam case

Hence, in fact there are three main actuator failure categories which two more common of them have been considered in the construction of the database: control restriction and surface jam, which are generally caused by physical damage, icing, or a loss of hydraulic power.

Table 2 presents the failure cases evaluated in the database. As can be seen, rudder failures have been chosen from lowest to highest degrees such that they cover different sections of the rudder operational range. Values in bracket are the lower limit ($LL$) and the upper limit ($UL$) of the rudder deflection, as in $[LL, UL]$. In fact, jamming failure is a special case of restriction failure in which $LL$ and $UL$ are identical. It should be noted that in the evaluation process of the impaired cases, the limits of the constraint (14) are changed to the $LL$ and $UL$ of the corresponding failure.

**Table 2 GTM-T2 control surface failures**

| Failure Type | Failure |
|---|---|
| Surface Jam | $-30°, -20°, -10°, 0°, 10°, 20°, 30°$ |
| Control Restriction | $[-30°, -20°], [-30°, -10°], [-30°, 0°], [-30°, 10°], [-30°, 20°], [20°, 30°], [10°, 30°], [0°, 30°],$ $[-10°, 30°], [-20°, 30°], [-20°, 20°],$ $[-20°, -10°], [-20°, 0°], [-20°, 10°], [-10°, 0°], [-10°, 10°], [10°, 20°], [0°, 20°], [-10°, 20°], [0°, 10°]$ |

In the employed database, MFEs of the unimpaired and impaired GTM have been evaluated at four different altitudes of Sea Level, 10000 ft, 20000 ft, and 30000 ft. It should be noted that the GTM-T2 is a sub-scale remotely-piloted air vehicle designed to fly at low altitudes only (line of sight), however, the MFEs were also evaluated at 20000 ft and 30000 ft using the GTM's MATLAB® – Simulink® model for the purpose of demonstrating and evaluating the envelope variations. Taking into account the unimpaired case and all considered rudder failure cases at the mentioned flying altitudes, the database is comprised of 3D MFEs of 112 different cases as shown in Table 3.

Table 4 presents the resolution increments in $V, \gamma$, and $\dot{\psi}$ ranges which were chosen such that high fidelity MFEs could be estimated. MFEs of the database were evaluated for different flight path angles within the range of $-5° \leq \gamma \leq 5°$. Hence, given the 1° resolution increment in the $\gamma$ range, each 3D MFE is composed of 11 different $\gamma$-constant 2D MFEs in the $(V - \dot{\psi})$ plane. Fig. 1 presents two instances of the evaluated 3D MFEs of the database.

**Table 3 Number of 3D maneuvering flight envelope**

| Failure Type | Impairments | Altitudes | Total |
|---|---|---|---|
| Unimpaired | N/A | 4 | 4 |
| Surface Jam | 7 | 4 | 28 |
| Control Restriction | 20 | 4 | 80 |
| Total | | | 112 |

**Table 4 Flight envelope increments**

| Parameter | Resolution Increment Size |
|---|---|
| $V$ | 1 knot |
| $\gamma$ | 1 deg |
| $\dot{\psi}$ | 0.2 deg/s |

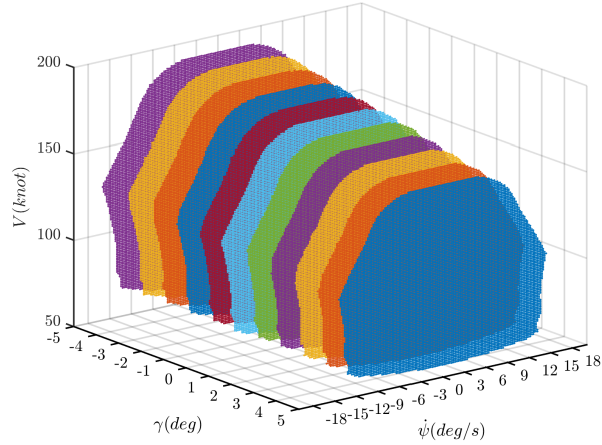 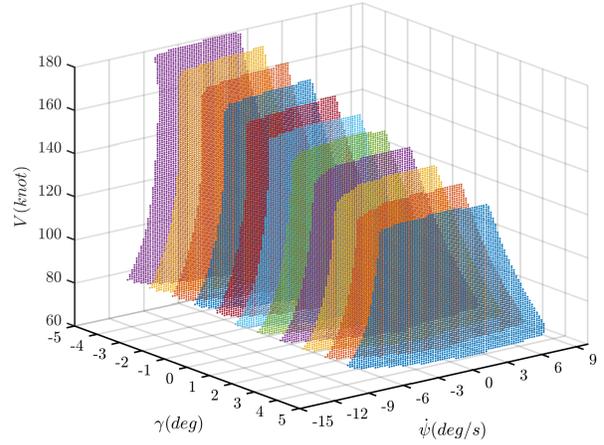

**a) Unimpaired case at sea level**     **b) Rudder jammed at 10° at 10000 ft**

**Fig. 1 Two instances of the Database MFEs utilized in this study**

## IV.  Approximating Models

In this section, linear and nonlinear least-squares methods are employed to fit approximating models to the data extracted from the database described in the previous section. As can be seen in Fig. 2, different flight conditions and various failure degrees result in completely different 2D MFEs in the $(V - \dot{\psi})$ plane. Specifically, it is shown that the centroid and the number of achievable trim points vary dramatically. Hence, as mentioned earlier, these two parameters which characterize the flight envelope variations are considered as the outputs of the approximating models.

Since trim points are characterized by four parameters $(h^*, V^*, \gamma^*, \dot{\psi}^*)$, altitude and flight path angle can be considered as two influencing parameters on the 2D MFEs. By increasing altitude flight envelope contracts due to the decrease in the air density and the engines' available thrust. Also, at higher flight path angles the required thrust is bigger due to the weight component $(W\gamma)$ along the velocity vector's axis; yielding in smaller flight envelopes. Hence, altitude and flight path angle are considered as two inputs to the models. Other inputs are identified based on the occurred failure type. In case of control surface failure, the new limitations on the surface deflection angle affect the 2D MFE. As mentioned in the previous section, both the jamming and control restriction failures can be characterized by the lower and upper limit values of the surface deflection angle. Therefore, the lower limit $(LL)$ and the upper limit $(UL)$ of the rudder deflection are also considered as two inputs to the approximating models. The aim is to develop analytical models which take the four inputs and accurately provide the intended outputs. These inputs and outputs are summarized in the Table 5.

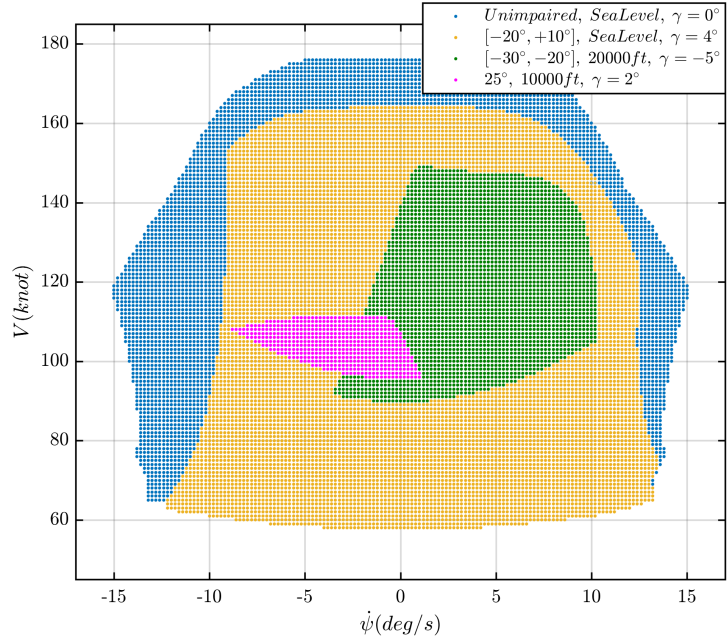

**Fig. 2 Variation of MFEs by different flight conditions and failure degrees**

**Table 5 Inputs and outputs of the approximating models**

| Inputs | Outputs |
| --- | --- |
| $h$ | |
| $\gamma$ | $n_{trim}$ (Number of trim points) |
| $LL$ | |
| $UL$ | $C_{trim}$ (Centroid of MFE) |

It should be noted that each of the two outputs presented in Table 5 is considered separately. In other words, in the following subsections, first, different methods are employed to develop approximating models capable of predicting the number of trim points at any a-priori unknown failure degree, and their results are compared. Then, the developed models which predict the centroid are presented and compared.

**A. Linear Least-Squares**

While a dataset does not explicitly describe the relationship between independent (predictor) and dependent (response or outcome) variables, regression can be used to fit an analytical parametric model to the dataset. One of the main types of regression is the least-squares data fitting method which estimates the model parameters or coefficients by minimizing the sum of squares of residuals or prediction errors which are the differences between the observed responses and the fitted model's outcomes for the input samples within the dataset.

$$minimize \quad \|\vec{e}\|^2 = \sum_{i=1}^{m} e_i^2 = \sum_{i=1}^{m}(y_i - \hat{y}_i)^2 = \sum_{i=1}^{m}(y_i - \hat{f}(\vec{\eta}, \vec{z}_i))^2 \quad (18)$$

While model $\hat{f}$ can be selected to be any arbitrary function, the problem is considered as linear regression if $\hat{f}$ is linear in the model parameters (coefficients) $\eta$ [51]

$$\hat{y} = \hat{f}(\vec{\eta}, \vec{z}) = \eta_1 \hat{f}_1(\vec{z}) + \cdots + \eta_{\mathcal{P}} \hat{f}_{\mathcal{P}}(\vec{z}) \quad (19)$$

This is equivalent to the model $\hat{f}$ being an affine function of the model parameter $\mathcal{P}$-vector $\eta$, as the first basis function $f_1$ is constant with value one. Hence, the $m$-vector of the prediction function's outcomes over the input samples can be written as

$$\vec{\hat{y}}_{m \times 1} = \mathcal{D}_{m \times \mathcal{P}} \, \vec{\eta}_{\mathcal{P} \times 1} \quad (20)$$

in which the design matrix $\mathcal{D}$ is defined as

$$\mathcal{D}_{ik} = \hat{f}_k(\vec{z}_i), \quad i = \{1, \cdots, m\}, \quad k = \{1, \cdots, \mathcal{P}\} \quad (21)$$

(20) also implies that the linear-in-the-coefficients requirement yields in the residual $e$ being an affine function of the model parameter $\eta$

$$\|\vec{e}_{m \times 1}\|^2 = \left\|\vec{\hat{y}}_{m \times 1} - \vec{y}_{m \times 1}\right\|^2 = \|\mathcal{D}_{m \times \mathcal{P}} \, \vec{\eta}_{\mathcal{P} \times 1} - \vec{y}_{m \times 1}\|^2 \quad (22)$$

Based on (18) and (21), the solution of the least-squares problem can be found by regressing the vector $\vec{y}_{m \times 1}$ onto the columns of $\mathcal{D}_{m \times \mathcal{P}}$ [51]. Assuming the columns of $\mathcal{D}$ to be linearly independent, the solution $\eta$ satisfies

$$\nabla[\|\mathcal{D}_{m \times \mathcal{P}} \, \vec{\eta}_{\mathcal{P} \times 1} - \vec{y}_{m \times 1}\|^2] = \partial[\|\mathcal{D}_{m \times \mathcal{P}} \, \vec{\eta}_{\mathcal{P} \times 1} - \vec{y}_{m \times 1}\|^2] \Big/ \partial \eta_k = 2\mathcal{D}^T(\mathcal{D}\vec{\eta} - \vec{y}) = 0 \quad k = 1, \cdots, \mathcal{P} \quad (23)$$

which can be written as

$$\mathcal{D}^T \mathcal{D} \vec{\eta} = \mathcal{D}^T \vec{y} \quad (24)$$

Since the Gram matrix $\mathcal{D}^T \mathcal{D}$ is invertible, the solution of the normal equations (24) can be obtained as

$$\vec{\eta} = (\mathcal{D}^T \mathcal{D})^{-1} \mathcal{D}^T \vec{y} = \mathcal{D}^\dagger \vec{y} \quad (25)$$

However, it should be noted that constructing the pseudo-inverse (Moore-Penrose inverse) $\mathcal{D}^\dagger$ requires forming the Gram matrix which is assumed to be nonsingular but in practice it could become singular due to rounding made in the floating point computations [51]. A more stable method to find the solution of (24) is the $QR$ factorization of $\mathcal{D}$ in which the design matrix $\mathcal{D}$ is expressed as the product of two matrices $Q$ and $R$ where $Q$ is an $m \times \mathcal{P}$ matrix with orthonormal columns and $R$ is a $\mathcal{P} \times \mathcal{P}$ upper triangular matrix with positive diagonal elements.

$$\vec{\eta} = \mathcal{D}^\dagger \vec{y} = (\mathcal{D}^T \mathcal{D})^{-1} \mathcal{D}^T \vec{y} = \left((QR)^T (QR)\right)^{-1} (QR)^T \vec{y} = R^{-1} Q^T \vec{y} \tag{26}$$

While the $QR$ factorization can be implemented through the Gram-Schmidt algorithm, a more reliable algorithm in the presence of round-off errors is the Housholder algorithm which is less sensitive to rounding error than the Gram-Schmidt algorithm [52].

The "linear in the coefficients" model presented in (19) resembles a polynomial with $p$ terms, where each term is a function of predictor variables. According to Table 5, $\vec{z} = [h \; \gamma \; LL \; UL]^T$, hence

$$n_{trim} = \sum_{j_1+j_2+j_3+j_4 \leq d} a_{j_1 j_2 j_3 j_4} h^{j_1} \gamma^{j_2} LL^{j_3} UL^{j_4} \tag{27}$$

where $a_{j_1 j_2 j_3 j_4}$ corresponds to the regression model coefficient $\eta_k$. Also $j_1$, $j_2$, $j_3$, and $j_4$ can be any non-negative integer as long as their sum is equal or less than the polynomial's total degree. It should be noted that $\vec{z}$ is also known as the predictor vector. All of the four predictor variables in $\vec{z}$ are autoscaled (i.e. scaled to unit variance) prior to the explained numerical procedure, which means each variable is multiplied by a scaling weight equal to the inverse of its standard deviation.

(27) is the general form of a $d$-degree polynomial approximating the number of trim points within an MFE from the provided input vector $\vec{z}$. Due to their continuous and differentiable nature, polynomials provide the possibility of defining and evaluating models based on analytical computation [53]. So far, polynomials have shown remarkable capabilities in modeling the GTM short period dynamics [54] and more recently in piecewise modeling of the full-envelope aerodynamic coefficients of the GTM [53, 55]. More specifically, the idea of interpolating flight envelopes in the offline database approach mentioned in Sec. I was corroborated by fitting first and second order polynomials to variation of the number of grid points inside reachable sets with flight condition or percentage of damage [33]. However, only limited cases were investigated and influencing parameters (i.e. flight condition or damage parameter) were taken into account independently. To the best of our knowledge, so far no related studies have modeled the aggregated effect of the main flight condition and failure-related factors on the flight envelope variations as a unified analytical model in the form of a polynomial.

In this research, the polynomial model of (27) has been developed with different degrees utilizing the described procedure to find the best approximating polynomial model possible (i.e. with the least prediction error). Specifically, a 2$^{nd}$ degree, a 3$^{rd}$ degree, and a 4$^{th}$ degree polynomial were built using the described database and their generalization abilities were compared. To have good generalization ability, the developed model should be able to predict the number of trim points for new unseen failure cases as well as it predicts the number of trim points on the failure cases used to form the model [51]. Therefore, employing the out-of-sample validation method, the database of the rudder failure cases was divided into two sets of training set and test set by a split ratio of 90%-10%. The polynomial models were fitted using only the data in the training set and their generalizations were assessed by their fit on the test set. Considering that the 112 3D MFEs presented in Table 3 are comprised of a total of 1102 2D MFEs, 111 (10%) 2D MFEs were selected randomly as the test set while the remaining 991 (90%) cases formed the training set.

Before presenting the results of the developed polynomial models, a few points should be mentioned: First, the models' generalizations were assessed by their Mean Squared Error (MSE) in the test set (i.e. $\|\vec{e}_{111\times1}\|^2/111$). The reason is that the generalization ability of the best developed polynomial will be compared with the neural network developed in the following subsections and usually the MSE is the main performance function used in developing, validating and testing of neural networks. Second, the evaluated MSEs are based on fitted outcomes and observed responses normalized between $-1$ and $1$. This is specifically important in the case of multi-element output ($C_{trim}$) as it ensures that the relative accuracy of $V$ and $\dot{\psi}$ which have different value ranges are treated as equally important instead of prioritizing the accuracy of $V$ over $\dot{\psi}$ due to larger value range. For the sake of integrity throughout the presented results, the same normalization has been applied to the case of ($n_{trim}$) even though it is a single-element output. Third, to better comprehend the generalization performance of the fitted models, in addition to providing the models' MSEs on the test set, the error percentages for five failure cases have also been calculated and provided as

$$err_{i=\{1,2,3,4,5\}} = [|y_i - \hat{y}_i|/y_i] \times 100\% \tag{28}$$

Whilst the first two of the five cases were selected from the test set, the other three cases are not even among the test data. In fact, these three cases were selected with input values that are not a combination of the considered altitudes and rudder deflection limits of the database. In other words, the altitudes and rudder upper and lower limits of these three cases have been selected as mid-range values to ensure that the developed models are not biased towards specific altitudes and failures. Tables 6 and 7 present the specifications and evaluation results of the constructed 2$^{nd}$, 3$^{rd}$, and 4$^{th}$ degree polynomial models.

**Table 6 Polynomial models' specifications and MSEs**

| Degree | Model | Number of coefficients | Adjusted $R^2$ value | Degrees of freedom (DOF) | Training set MSE | Test set MSE |
|---|---|---|---|---|---|---|
| 2 | Poly2222 | 15 | 0.9884 | 976 | 5.0251E-4 | 5.5858E-4 |
| 3 | Poly3333 | 35 | 0.9948 | 956 | 2.1851E-4 | 2.1058E-4 |
| 4 | Poly3344 | 68 | 0.9976 | 923 | 9.7601E-5 | 9.5569E-5 |

**Table 7 Polynomial models' error percentages for five failure cases**

| Failure case | $n_{trim}$ | Poly2222 | Poly3333 | Poly3344 |
|---|---|---|---|---|
| (1) | 4898 | 4483 (8.4645%) | 4756 (2.8975%) | 4852 (0.9436%) |
| (2) | 3759 | 3378 (10.1229%) | 3533 (6.0117%) | 3786 (0.7159%) |
| (3) | 3508 | 2847 (18.8338%) | 3109 (11.3663%) | 3285 (6.3497%) |
| (4) | 6538 | 6341 (3.0140%) | 6445 (1.4187%) | 6605 (1.0205%) |
| (5) | 4348 | 4063 (6.5507%) | 4157 (4.3960%) | 4208 (3.2164%) |

Even though the $R^2$ value indicates the amount of response's variation explained by the linear regression model, it could be misleading as it increases with every parameter added to the model regardless of whether the added term actually affects the model's response and the increased complexity is justified or not. Hence, the *adjusted $R^2$* which penalizes for the number of model terms is more appropriate for comparing polynomial models of different degrees.

As can be seen in Table 6, increasing the polynomial's total degree $d$ improves the model's ability to fit the variations and nonlinearities in the data. In the presented 2$^{nd}$ degree and 3$^{rd}$ degree models, each of the four variables' maximum degree equals the total degree of the polynomial; hence, the 2$^{nd}$ degree model can be shown as *Poly2222* whereas the 3$^{rd}$ degree model is *Poly3333*. Therefore, for these models the number of model parameters (i.e. polynomial terms) is $\binom{4+d}{d}$ as shown in Table 6. However, the 4$^{th}$ degree model (*Poly3344*) excludes the polynomial terms in which the maximum degrees of $h$ and $\gamma$ are 4, as the evaluated prediction errors for the aforementioned five failure cases indicate that the *Poly4444* model is over – specified and for a number of test cases the model outcomes are largely erroneous. Based on the obtained results, the *Poly3344* is properly specified and resolves the overfitting problem of the *Poly4444* model. Detailed results of the *Poly4444* model and more discussion on this are presented in the Appendix A.

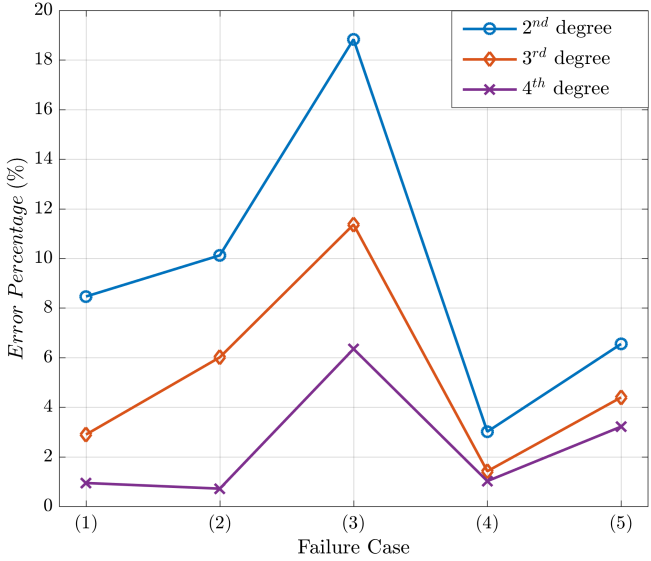

Fig. 3 Polynomial models' prediction error percentages for the five test cases of Table 7

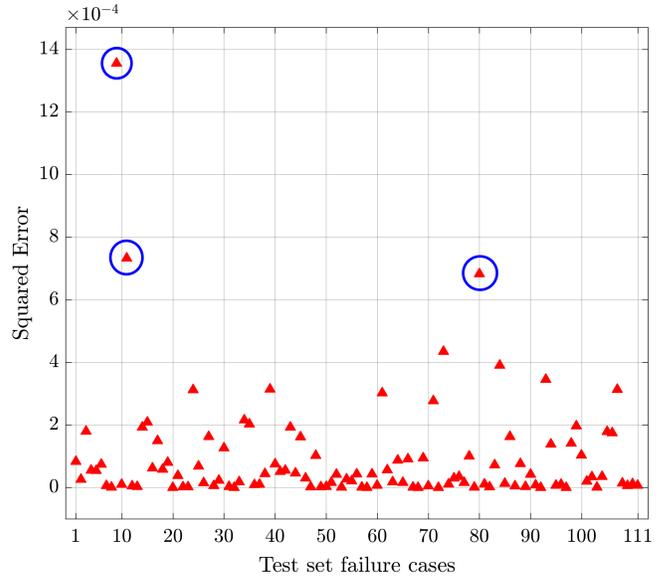

Fig. 4 Squared errors in predicting $n_{trim}$ over the test set (Poly3344)

Figure 3 shows the five failure cases' prediction error percentages presented in Table 7 for better comprehension. Also, analyzing the squared errors for the 111 failure cases of the test set as presented in Fig. 4 enables identifying the regions of the training set where more data samples are needed. It is shown in Fig. 4 that for most of the 111 cases the squared error value of the *poly3344* is in the order of $10^{-5}$ yielding a test set MSE of 9.5569E-5. However, the blue-circled cases have relatively higher errors than the rest of the test cases. Since all the three cases belong to the same altitude (i.e. 30000 ft), it suggests that if needed; the MSE of the model can be further improved by adding more training cases at this altitude.

Since the subject model is five dimensional ($n_{trim} = f(h, \gamma, LL, UL)$) it can be visualized in 3D by fixing any two out of the four input variables. Figures 6, 7, and 9 each present an instance of these 3D visualizations along with the corresponding training data and test data as red and black dots, respectively. For instance, Fig. 7 presents the fitted 4$^{th}$ degree polynomial model in which the *LL* has been set equal to the *UL* and $\gamma = 2°$. In other words, the figure depicts the variation of the number of trim points $n_{trim}$ with various altitudes and different jamming angles at $\gamma = 2°$. Also, Fig. 8 displays the prediction bounds for the fitted function with 95% confidence level. Both figures indicate good fit and proper generalization of the *poly3344* model. It should be noted that figures 6 to 9 are just four examples among several possible 3D visualizations intended to depict the *poly3344* model as fitted surfaces along with training and test data points for better comprehension. Hence, they only represent portions of the training and test data. However, the accuracy of the *poly3344* model over the entire dataset; besides being demonstrated by the numerical results presented in the tables 6 and 7, is also evident in Fig. 5 which represents the random scattering of the residuals around zero for all the 991 training cases and 111 test cases.

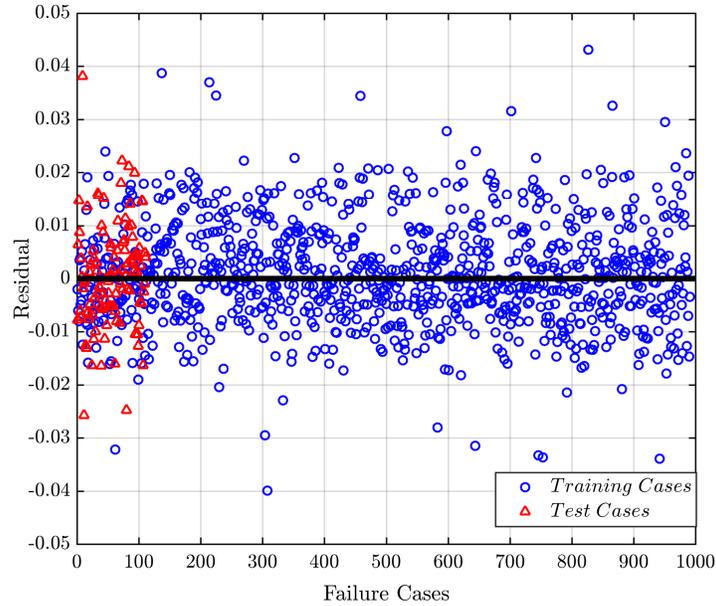

**Fig. 5 $n_{trim}$ prediction residuals for the 991 training cases and 111 test cases (Poly3344)**

According to the presented results and accomplished investigation of the Appendix A, the fitted *poly3344* model is the best linear regression model as consideration of any model with higher degrees in any of the four input variables yields in an ill-conditioned Vandermonde design matrix $\mathcal{D}$.

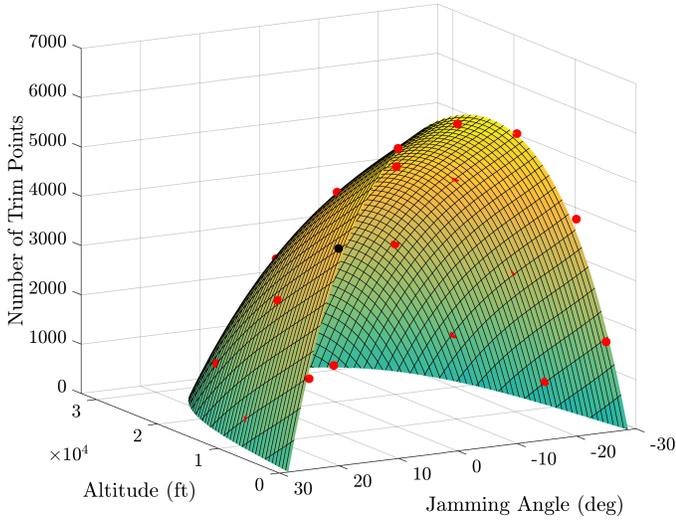

**Fig. 6 3D visualization of the 4th degree polynomial model at $\gamma = 2°$ (Poly3344)**

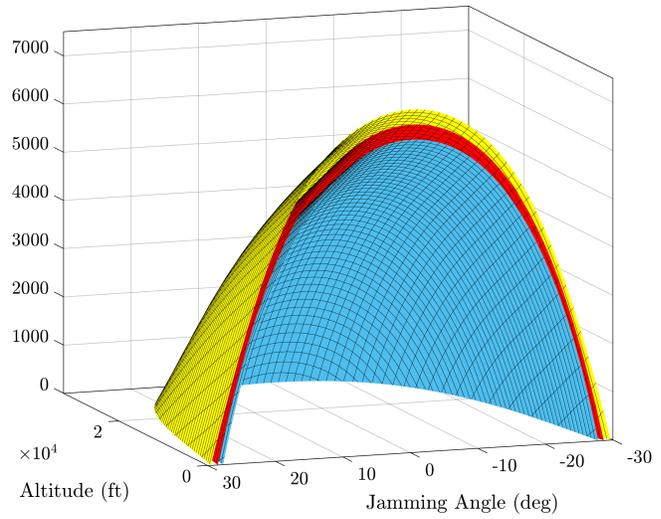

**Fig. 8 3D visualization of the 95% confidence prediction bounds for the Poly3344 at $\gamma = 2°$**

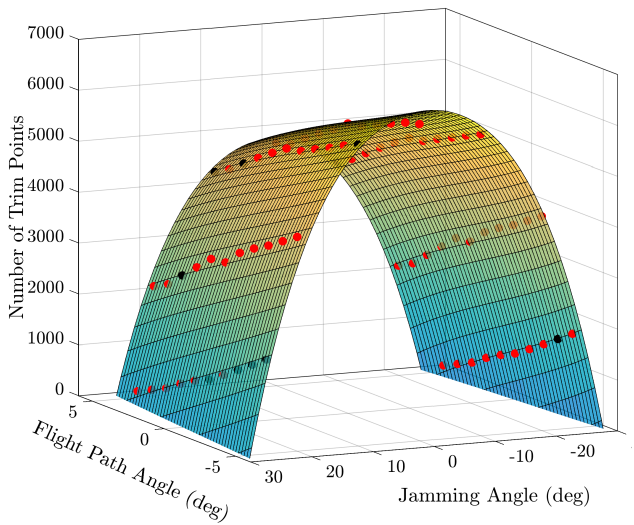

**Fig. 7 3D visualization of the 4th degree polynomial model at $h = 10000$ ft (Poly3344)**

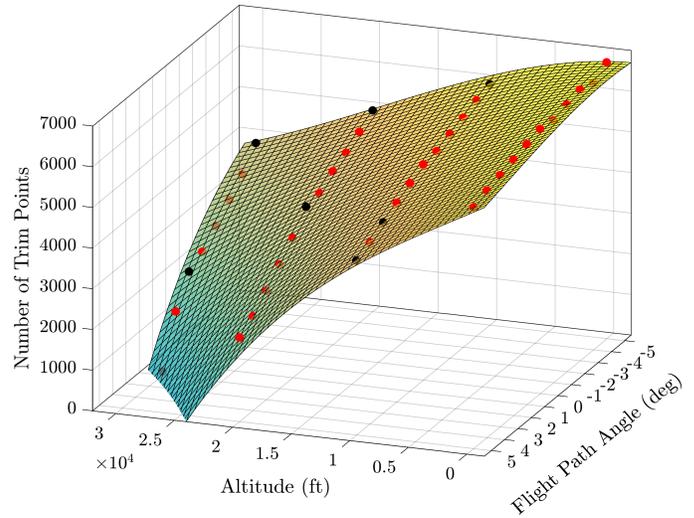

**Fig. 9 3D visualization of the 4th degree polynomial model at 10° jamming angle (Poly3344)**

### B. Nonlinear Least-Squares

Alternatively, if the model $\hat{f}$ in the data fitting problem of (18) is selected to be an arbitrary equation which is nonlinear or not entirely linear in the coefficients, then the nonlinear least-squares method is used to estimate the model parameters.

$$minimize \ \ F(\vec{\eta}) = \|\vec{e}\|^2 = \sum_{i=1}^{m} e_i^2 = \sum_{i=1}^{m}(y_i - \hat{y}_i)^2 = \sum_{i=1}^{m}(y_i - \hat{f}(\vec{\eta}, \vec{z}_i))^2 \tag{29}$$

Since the objective function $F(\vec{\eta})$ is nonlinear in the model parameters, a single pass fitting procedure cannot be used to obtain the model coefficients and an iterative scheme is needed. Two main algorithms widely used to tackle the unconstrained optimization problem of (29) are Trust-Region-Reflective and Levenberg-Marquardt algorithms. Each algorithm is briefly elaborated in the following subsections.

1. *Trust-Region-Reflective algorithm*

The basic idea of the Trust-Region method is to approximate the objective function $F$ with a simpler function that reflects the behavior of the function $F$ in a neighborhood (i.e. trust region) $O$ around the current point $\eta$ and to minimize the approximating function over $O$ to find a trial step $\rho$. The approximating function is defined to be the first two terms of the Taylor series of the objective function $F$. Hence, (29) is stated in the form of a quadratic minimization as the Trust-Region subproblem

$$minimize \ \ \{g^T \vec{\rho} + \frac{1}{2} \vec{\rho}^T \mathbb{H} \vec{\rho} : \|\vec{\rho}\| \leq \Delta\} \tag{30}$$

Once the step $\rho$ is determined the current point $\eta$ is updated to $\eta + \rho$ if $F(\vec{\eta} + \vec{\rho}) < F(\vec{\eta})$ and the Trust-Region dimension $\Delta$ is adjusted, otherwise the current point remains unchanged, the trust region $O$ is contracted and the step $\rho$ is recalculated. Considering that $\eta$ is a $\mathcal{P}$-vector, $\rho$ belongs to a $\mathcal{P}$-space and $g \in R^{\mathcal{P}}, \mathbb{H} \in R^{\mathcal{P} \times \mathcal{P}}$. Hence, minimizing the quadratic function of (30) over the entire $p$-space is time consuming and computationally expensive. An approximation approach is to restrict the Trust-Region subproblem to a subspace spanned by two reasonably chosen directions [56].

$$minimize \ \ \{g^T \vec{\rho} + \frac{1}{2} \vec{\rho}^T \mathbb{H} \vec{\rho} : \|\vec{\rho}\| \leq \Delta, \ \rho \in [\vec{\rho}_1, \vec{\rho}_2]\} \tag{31}$$

While generally $\rho_1$ and $\rho_2$ are selected to be in the direction of the gradient of the objective function and an approximate Newton direction respectively if $\mathbb{H}$ is positive definite

$$\vec{\rho}_1 = -g \ , \ \vec{\rho}_2 = -\mathbb{H}^{-1} g \tag{32}$$

in case of the nonlinear least-squares problem, $\rho_2$ is defined as an approximate Gauss-Newton direction which does not require the second derivatives of the components $e_i$ to be computed

$$min \ \ \|\mathbb{J} \vec{\rho}_2 + \vec{e}\|^2 \tag{33}$$

and is obtained by approximately solving the following normal equations using the preconditioned conjugate gradients method [57].

$$\mathbb{J}^T\mathbb{J}\vec{\rho}_2 = -\mathbb{J}^T\vec{e} \tag{34}$$

The two-dimensional minimization of (31) is performed at each iteration until convergence.

2. *Levenberg-Marquardt algorithm*

The Levenberg-Marquardt algorithm is a variation of Newton's method which has been designed for optimization of the sum of squares of nonlinear functions. Within Newton's method, the search direction for the problem presented in (29) is calculated at the $j^{th}$ iteration as

$$\mathfrak{d}_j = -\mathbb{H}(\vec{\eta}|_j)^{-1}\mathbb{g}(\vec{\eta}|_j) \tag{35}$$

where

$$[\mathbb{g}(\vec{\eta})]_{k=\{1,2,\ldots,\mathcal{P}\}} = [\nabla F(\vec{\eta})]_k = 2\sum_{i=1}^{m} e_i(\vec{\eta}) \left.\partial e_i(\vec{\eta})\right/\partial \eta_k \Rightarrow \mathbb{g}(\vec{\eta}) = \nabla F(\vec{\eta}) = 2\mathbb{J}^T(\vec{\eta})e(\vec{\eta}) \tag{36}$$

$$[\mathbb{H}(\vec{\eta})]_{k,j=\{1,2,\ldots,\mathcal{P}\}} = [\nabla^2 F(\vec{\eta})]_{k,j} = 2\sum_{i=1}^{m}\{\frac{\partial e_i(\vec{\eta})}{\partial \eta_k}\frac{\partial e_i(\vec{\eta})}{\partial \eta_j} + e_i(\vec{\eta})\frac{\partial^2 e_i(\vec{\eta})}{\partial \eta_k \partial \eta_j}\} \Rightarrow \mathbb{H}(\vec{\eta}) = \nabla^2 F(\vec{\eta}) \tag{37}$$

$$= 2\mathbb{J}^T(\vec{\eta})\mathbb{J}(\vec{\eta}) + 2M(\vec{\eta})$$

The matrix $M(\vec{\eta}) = e_i(\vec{\eta})\nabla^2 e_i(\vec{\eta})$ tends to zero as so does the residual $e_i$ when $\eta$ approaches the solution. Thus, assuming that $M(\vec{\eta})$ is small and can be ignored, the Hessian matrix would become $\nabla^2 F(\vec{\eta}) \cong 2\mathbb{J}^T(\vec{\eta})\mathbb{J}(\vec{\eta})$. So, the search direction can be rewritten as

$$\mathfrak{d}_j = \vec{\eta}|_{j+1} - \vec{\eta}|_j = -[\mathbb{J}^T(\vec{\eta}|_j)\mathbb{J}(\vec{\eta}|_j)]^{-1}\mathbb{J}^T(\vec{\eta}|_j)e(\vec{\eta}|_j) \tag{38}$$

(38) which does not require the calculation of second derivatives is the Gauss-Newton method [58]. However, if $M(\vec{\eta})$ is not small enough it cannot be omitted and even if it is small enough, the $\mathbb{J}^T(\vec{\eta})\mathbb{J}(\vec{\eta})$ matrix may be non-invertible. In both cases the Gauss-Newton method encounters problem which can be overcome by a modified Hessian matrix as in the Levenberg-Marquardt method [59].

$$\mathbb{H}(\vec{\eta}) = \mathbb{J}^T(\vec{\eta})\mathbb{J}(\vec{\eta}) + \xi\mathbb{I} \tag{39}$$

which subsequently yields in a modified search direction

$$\mathfrak{d}_{\hat{j}} = \vec{\eta}|_{\hat{j}+1} - \vec{\eta}|_{\hat{j}} = -[\mathbb{J}^T(\vec{\eta}|_{\hat{j}})\mathbb{J}(\vec{\eta}|_{\hat{j}}) + \xi\mathbb{I}]^{-1}\mathbb{J}^T(\vec{\eta}|_{\hat{j}})e(\vec{\eta}|_{\hat{j}}) \qquad (40)$$

The algorithm initializes with $\xi = 0.01$. If an iteration yields in a smaller objective function value, then $\xi$ is divided by some factor so that the algorithm would approach the Gauss-Newton method, otherwise, $\xi$ is multiplied by some factor so that the search direction tends towards the steepest descent direction which eventually results in a smaller objective function value. Hence, the Levenberg-Marquardt algorithm is a compromise between the guaranteed convergence of the steepest descent and the convergence speed of the Gauss-Newton method [58].

Generally, both the Trust-Region-Reflective and Levenberg-Marquardt algorithms are started with random values as the initial estimate of the model coefficients. Moreover, in both algorithms $\mathbb{J}$ is numerically approximated using the finite differences which are computationally expensive for problems with many model parameters.

*3. Objective Function*

In this research, the nonlinear model $\hat{f}$ in the objective function $F(\vec{\eta})$ is defined as a linear function of another function $\mathcal{T}$:

$$F(\vec{\eta}) = \sum_{i=1}^{m}(y_i - \hat{f}(\vec{\eta}, \vec{z}_i))^2 \qquad (41)$$

$$\hat{f}(\vec{\eta}, \vec{z}) = \eta_1 + \eta_2\,\mathcal{T}(\vec{\eta}, \vec{z}) \qquad (42)$$

and $\mathcal{T}$ is the hyperbolic tangent function

$$\mathcal{T}(\vec{\eta}, \vec{z}) = \left[2 \big/ \left[1 + e^{-2(\eta_3 h + \eta_4 \gamma + \eta_5 LL + \eta_6 UL + \eta_7)}\right]\right] - 1 \qquad (43)$$

The hyperbolic tangent function is an S-shaped rescaling of the logistic sigmoid function that maps the resulting values between $-1$ and $1$. Both functions have shown great capabilities in learning the nonlinearities between the inputs and outputs and have been widely used in nonlinear regression. However, the hyperbolic tangent function has the advantage over logistic sigmoid function that instead of strongly negative inputs being mapped to near zero values, they are mapped to negative values while only zero inputs being mapped to near zero values. Therefore, the hyperbolic tangent function is used in this research. Furthermore, the model $\hat{f}$ of (42) is set to be a linear function of the hyperbolic tangent function $\mathcal{T}$. This linear function addresses the extrapolation effect which is a major issue in regression problems. Instead of extrapolating by projecting the trend of the fitted model onwards, the mean output value of the available data could be considered as the extrapolation result. However, the sigmoidal functions saturate at either the minimum or maximum, but a linear function does not saturate and so can extrapolate further [60].

The nonlinear model of (42) is comprised of 7 coefficients. Similar to the polynomial models fitted by the linear least-squares method, it is expected that increasing the number of coefficients of the nonlinear-in-the-parameters model $\hat{f}$ would result in more complex models that better resemble the nonlinear relationships between the inputs and outputs of the dataset. Therefore, two other models with 13 and 19 coefficients are also considered and their goodness of fit and generalizations are evaluated.

$$\hat{f}(\vec{\eta}, \vec{z}) = \eta_1 + \eta_2 \left( \left[ 2 \big/ [1 + e^{-2(\eta_3 h + \eta_4 \gamma + \eta_5 LL + \eta_6 UL + \eta_7)}] \right] - 1 \right) +$$
$$\eta_8 \left( \left[ 2 \big/ [1 + e^{-2(\eta_9 h + \eta_{10} \gamma + \eta_{11} LL + \eta_{12} UL + \eta_{13})}] \right] - 1 \right) \tag{44}$$

$$\hat{f}(\vec{\eta}, \vec{z}) = \eta_1 + \eta_2 \left( \left[ 2 \big/ [1 + e^{-2(\eta_3 h + \eta_4 \gamma + \eta_5 LL + \eta_6 UL + \eta_7)}] \right] - 1 \right) +$$
$$\eta_8 \left( \left[ 2 \big/ [1 + e^{-2(\eta_9 h + \eta_{10} \gamma + \eta_{11} LL + \eta_{12} UL + \eta_{13})}] \right] - 1 \right) + \eta_{14} \left( \left[ 2 \big/ [1 + e^{-2(\eta_{15} h + \eta_{16} \gamma + \eta_{17} LL + \eta_{18} UL + \eta_{19})}] \right] - 1 \right) \tag{45}$$

As can be seen, the 13-coefficient and 19-coefficient models ($\hat{f}_{13}, \hat{f}_{19}$) are extensions of the 7-coefficient model ($\hat{f}_7$) constructed by adding more hyperbolic tangent functions:

$$\hat{f}_7 = b_1 + a_1 \mathcal{T}_1 \quad , \quad \hat{f}_{13} = b_1 + a_1 \mathcal{T}_1 + a_2 \mathcal{T}_2 \quad , \quad \hat{f}_{19} = b_1 + a_1 \mathcal{T}_1 + a_2 \mathcal{T}_2 + a_3 \mathcal{T}_3 \tag{46}$$

It should be noted that after utilizing the explained algorithms in several instances of optimizing the objective function of (41) formed by the nonlinear model of (42), it was found that both algorithms find nearly the same solution starting from the same initial point, however, the Levenberg-Marquardt (LM) converges with fewer iterations and function evaluations than the Trust-Region-Reflective (TRR). Therefore the Levenberg-Marquardt algorithm, which appears to be more efficient, was selected to fit the nonlinear models ($\hat{f}_{13}, \hat{f}_{19}$). The following tables present the obtained results for the nonlinear least-squares data fitting problem. For the purpose of comparison, the same training set and test set used with the linear least-squares method were also utilized for all the nonlinear regressions of this section.

**Table 8 Comparison of nonlinear regression models' specifications**

| Model | Algorithm | Number of iterations | Number of function evaluations | First-order optimality | Step size | Squared 2-norm of residual | Test set MSE |
|---|---|---|---|---|---|---|---|
| $\hat{f}_7$ | TRR | 20 | 168 | 0.0129 | 0.0045 | 33.3560 | 0.0320 |
| $\hat{f}_7$ | LM | 13 | 119 | 0.0050 | 0.0020 | 33.3560 | 0.0320 |
| $\hat{f}_{13}$ | LM | 88 | 1305 | 0.0772 | 0.0459 | 16.515 | 0.0149 |
| $\hat{f}_{19}$ | LM | 391 | 8036 | 0.0481 | 0.0311 | 1.9424 | 0.0028 |

**Table 9 LM-trained models' error percentages for the five failure cases**

| Failure case | $n_{trim}$ | $\hat{f}_7$ | $\hat{f}_{13}$ | $\hat{f}_{19}$ |
|---|---|---|---|---|
| (1) | 4898 | 3565 | 4419 | 4942 |
|     |      | (27.2205 %) | (9.8747%) | (0.9075%) |
| (2) | 3759 | 2959 | 3374 | 3608 |
|     |      | (21.2755%) | (10.2517%) | (4.0234%) |
| (3) | 3508 | 2700 | 2825 | 3213 |
|     |      | (23.0414%) | (19.4566%) | (8.4201%) |
| (4) | 6538 | 5682 | 6028 | 6519 |
|     |      | (13.0917%) | (7.8032%) | (0.2944%) |
| (5) | 4348 | 2579 | 3244 | 4141 |
|     |      | (40.6875%) | (25.3869%) | (4.7507%) |

It should be noted that different initial estimates of the model parameters lead to different results as the optimization might get trapped in different local minimas. Therefore, in order to make the results comparable, the same initial values were used for all of the four optimizations of Table 8. To do so, a 19-vector $\eta_0$ was generated randomly and the excess parameters were omitted for the models with lower number of coefficients. Also, the nonlinear least-squares data fitting processes were executed several times starting from different random $\eta_0$ vectors and the best obtained results are presented here.

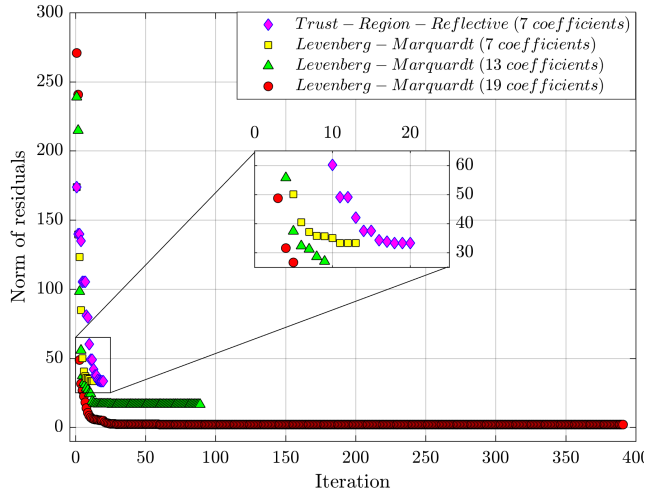

**Fig. 10 Norms of residuals of the 4 nonlinear regression models presented in Table 10**

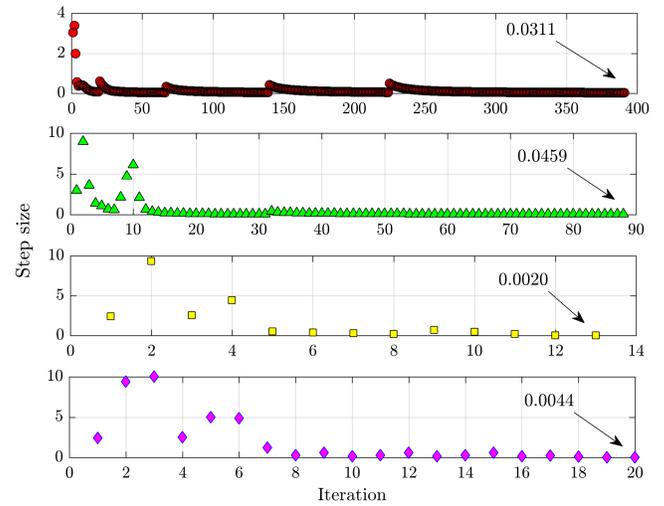

**Fig. 11 Step sizes of the 4 nonlinear regression models presented in Table 10**

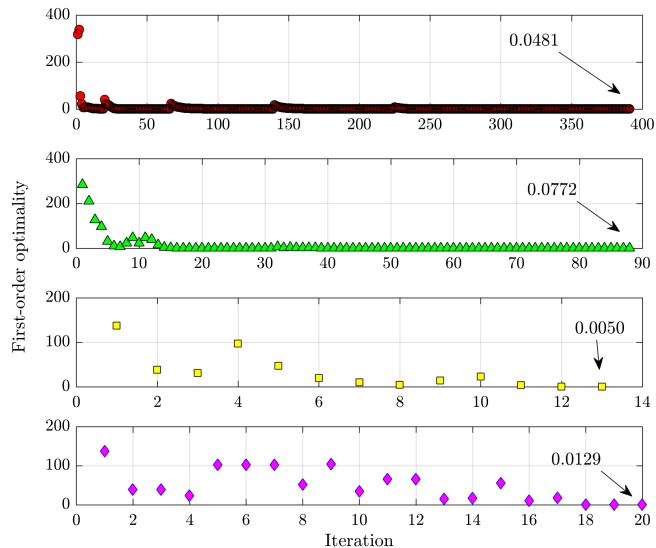

**Fig. 12 First-order optimality of the 4 nonlinear regression models presented in Table 10**

As can be seen in figures 10 to 12 and Table 8, increasing model parameters has led to better performance. Both the first-order optimality and the step size are measures of how close the estimated model parameters are to their optimal values. The obtained values of these measures show that all the optimization processes have almost reached to the optimum. However, the differences in the squared norm of residual (error) values reflect the power of more complex models in adapting to the training data. The test set MSEs and the error percentages of the five failure cases reveal the same conclusion for the unseen data. In other words, Table 8 presents the best fit and generalization possible with the corresponding number of coefficients. As the best obtained results (which belong to the 19-coefficient model) are still behind those obtained by the best fitted polynomial, models with higher number of parameters must be evaluated.

*4. Nonlinear regression by neural networks*

As will be discussed in the following, it is more effective to carry out the nonlinear parameters estimation process (i.e. nonlinear regression) of models with higher number of coefficients within the neural network training framework. In fact, feedforward networks used for function fitting often have one or more hidden layers with logistic-sigmoid or tangent-sigmoid transfer function followed by an output layer with linear transfer function [61]. Specifically, the hyperbolic tangent function $\mathcal{T}$ in (43) resembles the tangent-sigmoid hidden neuron of a multilayer perceptron, and the linear function of (42) is analogous to the linear output neuron.

Fig. 13 shows the architecture of the aforementioned two-layer network which as a universal approximator is capable of approximating any nonlinear function provided that there are sufficient neurons in the hidden layer [62]. In the presented architecture, $R$ and $S$ indicate the number of parameters in the input layer and the number of neurons in the corresponding layer, respectively. Also, $a^1$, $a^2$, $W^1$, and $W^2$ represent the outputs of the tan-sigmoid transfer function and the linear transfer function, and the matrices of the hidden layer's weights and the output layer's weights. The fitted models $\hat{f}_7$, $\hat{f}_{13}$, and $\hat{f}_{19}$ correspond to three neural networks of the presented architecture with one, two, and three hidden neurons, respectively. In these networks $R$ equals four and $S^2 = 1$ when the output is $(n_{trim})$ and $S^2 = 2$ in the case of $(C_{trim})$.

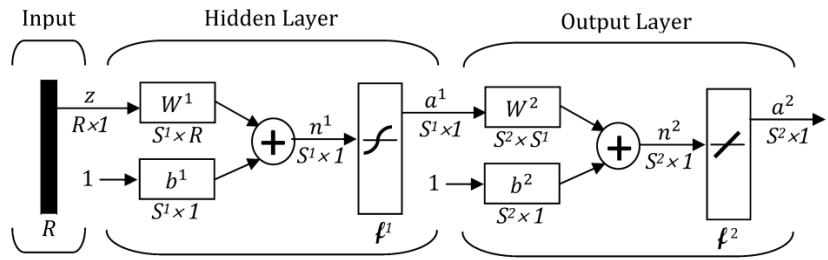

**Fig. 13 Two-layer feedforward neural network architecture**

The reason that the considered nonlinear regression can be implemented more effectively by training neural networks with the Levenberg-Marquardt algorithm is twofold. First, a modification of the backpropagation algorithm is incorporated into the Levenberg-Marquardt algorithm for computation of the Jacobian matrix $\mathbb{J}$ in (40). Backpropagation which is a training method for neural networks is an approximate steepest descent algorithm in which the objective function is mean squared error which is proportional to the sum of squared errors (i.e. the objective function in (29)) assuming the targets of the dataset occur with equal probabilities [58].

In the standard backpropagation algorithm, the objective function is approximated by replacing the sum of squared errors with the squared error of a single input-output pair of the training set. Then the input to the network is propagated forward until the output of the last layer (i.e. $a^2$ in the presented architecture) is calculated. Finally, the weights and biases $(W, b)$ of the network (i.e. the model coefficients) are updated at each iteration by evaluating the derivatives of the objective function with respect to each layer's weights and biases, which themselves are proportional to the sensitivity of the objective function to changes in the net input of the $\mathcal{L}^{th}$ neuron in that layer. As each layer's sensitivity is a function of the sensitivity in the next layer, it is computed through a recurrence relationship which propagates back the sensitivities from the last layer to the first layer [58]. Thus, for the presented architecture of Fig. 13:

$$W^{\mathbb{c}}_{\mathcal{L},\mathcal{J}}(\mathbb{k}+1) = W^{\mathbb{c}}_{\mathcal{L},\mathcal{J}}(\mathbb{k}) - \sigma \partial F / \partial W^{\mathbb{c}}_{\mathcal{L},\mathcal{J}} = W^{\mathbb{c}}_{\mathcal{L},\mathcal{J}}(\mathbb{k}) - \sigma \partial F / \partial n^{\mathbb{c}}_{\mathcal{L}} \partial n^{\mathbb{c}}_{\mathcal{L}} / \partial W^{\mathbb{c}}_{\mathcal{L},\mathcal{J}} = W^{\mathbb{c}}_{\mathcal{L},\mathcal{J}}(\mathbb{k}) - \sigma \lambda^{\mathbb{c}}_{\mathcal{L}} a^{\mathbb{c}-1}_{\mathcal{J}} \quad (47)$$

$$b^{\mathbb{c}}_{\mathcal{L}}(\mathbb{k}+1) = b^{\mathbb{c}}_{\mathcal{L}}(\mathbb{k}) - \sigma \partial F / \partial b^{\mathbb{c}}_{\mathcal{L}} = b^{\mathbb{c}}_{\mathcal{L}}(\mathbb{k}) - \sigma \partial F / \partial n^{\mathbb{c}}_{\mathcal{L}} \partial n^{\mathbb{c}}_{\mathcal{L}} / \partial b^{\mathbb{c}}_{\mathcal{L}} = b^{\mathbb{c}}_{\mathcal{L}}(\mathbb{k}) - \sigma \lambda^{\mathbb{c}}_{\mathcal{L}} \quad (48)$$

where $\sigma$ is the learning rate and,

$$\mathbb{c} = \{1,2\}, \mathcal{L} = [\{1,\dots,S^1\} \; or \; \{1,\dots,S^2\}], \mathcal{J} = [\{1,\dots,R\} \; or \; \{1,\dots,S^1\}],$$

$$\lambda^{\mathbb{c}} = \mathbb{M}^{\mathbb{c}}(W^{\mathbb{c}+1})^T \lambda^{\mathbb{c}+1}, \quad \lambda^2 = -2\mathbb{M}^2 e_i, \quad i = \{1,\dots,m\} \quad (49)$$

and $\mathbb{M}^{\mathbb{c}}$ is a diagonal matrix comprising the derivatives of each layer's neurons' outputs to that layer's neurons' inputs $\partial a^{\mathbb{c}}_{\mathcal{L}} / \partial n^{\mathbb{c}}_{\mathcal{L}}$.

The terms in the Jacobian matrix $\mathbb{J}$ that need to be computed are $\partial e_i(\vec{\eta})/\partial \eta_k$. Considering the contributions of each of the last layer's neurons to the error of a single input-output pair ($e_i$) and also that the model parameters $\eta$ are the network's weights and biases, the terms in $\mathbb{J}$ can be rewritten as

$$\partial e_{\mathfrak{h}} / \partial \eta_k = \partial e_{\varepsilon,i} / \partial W^{\mathbb{c}}_{\mathcal{L},\mathcal{J}} \left( or \; \partial e_{\varepsilon,i} / \partial b^{\mathbb{c}}_{\mathcal{L}} \right) = \tilde{\lambda}^{\mathbb{c}}_{\mathcal{L},\mathfrak{h}} a^{\mathbb{c}-1}_{\mathcal{J},i} \left( or \; \tilde{\lambda}^{\mathbb{c}}_{\mathcal{L},\mathfrak{h}} \right) \quad (50)$$

$$\mathfrak{h} = (i-1)S^2 + \varepsilon, \qquad i = \{1, \ldots, m\}, \qquad \varepsilon = \{1, \ldots, S^2\}$$

where $\tilde{\lambda}$ is the Marquardt sensitivity defined in a similar way that the sensitivity $\lambda$ in the standard backpropagation method is defined. Furthermore, the last layer's sensitivity is modified such that $\tilde{\lambda}^2 = -\mathbb{M}^2$ [58].

Another advantage of nonlinear regression by training the presented two-layer feedforward neural network is that the initial values of the model parameters (i.e. weights and biases) can be generated more intelligently such that the training time is reduced, rather than being generated purely random as in the optimization problem presented previously. Specifically, the Nguyen-Widrow method [63] generates initial weight and bias values for a layer such that the active regions of the layer's neurons are distributed approximately evenly across the layer's input space. While pure randomly selected weights change during the training process until the region of interest is divided into small intervals, pre-selecting their values such that each neuron is assigned its own interval at the start of the training would eliminate the majority of the weight variations and hence will yield in less training time. It should be noted that the initial values selected with this method still contain a degree of randomness as first they are selected from a uniform random distribution and then their magnitudes are adjusted accordingly. This feature results in different weights' and biases' initial values each time the same network structure is trained, which is useful to find the best network from multiple trainings. Technical details of the Nguyen-Widrow implementation can be found in [63].

As mentioned previously, the fitted models $\hat{f}_7$, $\hat{f}_{13}$, and $\hat{f}_{19}$ correspond to three feedforward neural networks of the presented two-layer architecture with one, two, and three hidden neurons, and their results suggest that increasing the hidden layer neurons (i.e. the number of model parameters) would yield in a better fit. Therefore, several networks with different sizes were built by increasing the number of hidden neurons one by one. Each network was trained 15 times (starting from different weights and biases initial values) using the same training set of the previous methods and evaluated each time on the same test set used for the polynomials. The network with the least MSE on the test set was selected among the 15 trained networks as the best generalizing network with the corresponding number of hidden neurons. After comparing the best networks of different sizes it was found that up to 10 hidden neurons the test MSE is reduced by increasing the network size, however, adding more hidden neurons unreasonably increases the network size and consequently the training time and the chance for the network to overfit the data, as it does not improve the network generalization anymore. Hence, the best network with 10 hidden neurons was selected as the final network modeling the nonlinear relationship between the input parameters $[h\ \gamma\ LL\ UL]$ and the output ($n_{trim}$).

**Table 10 Best neural network's specifications and MSEs against *Poly3344***

| Model | Number of coefficients | Adjusted $R^2$ value | Training set MSE | Test set MSE | Total fitting run time (sec) |
|---|---|---|---|---|---|
| 2-layer 10-hidden neuron Neural Network | 61 | 0.9995 | 4.0120E-5 | 4.0221E-5 | 10.865 |
| Poly3344 | 68 | 0.9976 | 9.7601E-5 | 9.5569E-5 | 0.062 |

**Table 11 Best neural network's error percentages for the five failure cases against *Poly3344***

| Failure case | $n_{trim}$ | Poly3344 | 2-layer 10-hidden neuron Neural Network |
|---|---|---|---|
| (1) | 4898 | 4852 (0.9436%) | 4876 (0.4455%) |
| (2) | 3759 | 3786 (0.7159%) | 3738 (0.5471%) |
| (3) | 3508 | 3285 (6.3497%) | 3556 (1.3852%) |
| (4) | 6538 | 6605 (1.0205%) | 6434 (1.5925%) |
| (5) | 4348 | 4208 (3.2164%) | 4224 (2.8431%) |

Tables 10 and 11 present the obtained results of the final network in comparison with those obtained with the best model developed by the linear least-squares method. Also, figures 14 to 16 present the training performance, error histogram, and the regression plots of the final network, respectively. It should be noted that during the training process of neural networks, a small portion of the training set is used as validation data. In this research, the whole available dataset is divided by into training data (80%), validation data (10%), and test data (10%). The test data is the same test set used previously, so the obtained results of the neural networks are comparable with the previous methods.

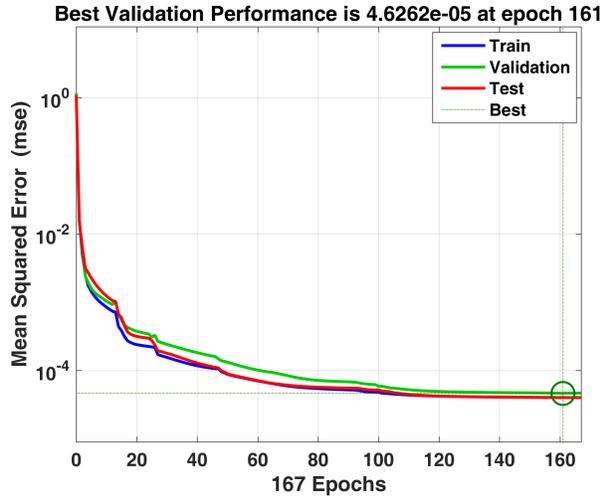

Fig. 14 Training performance of the best neural network for ($n_{trim}$) case

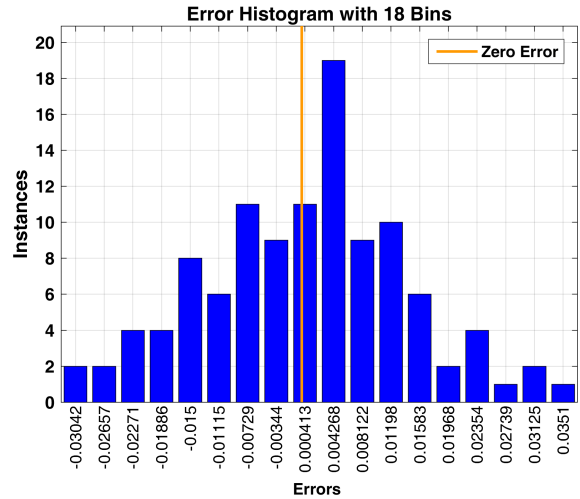

Fig. 15 Error histogram of the best neural network for ($n_{trim}$) case

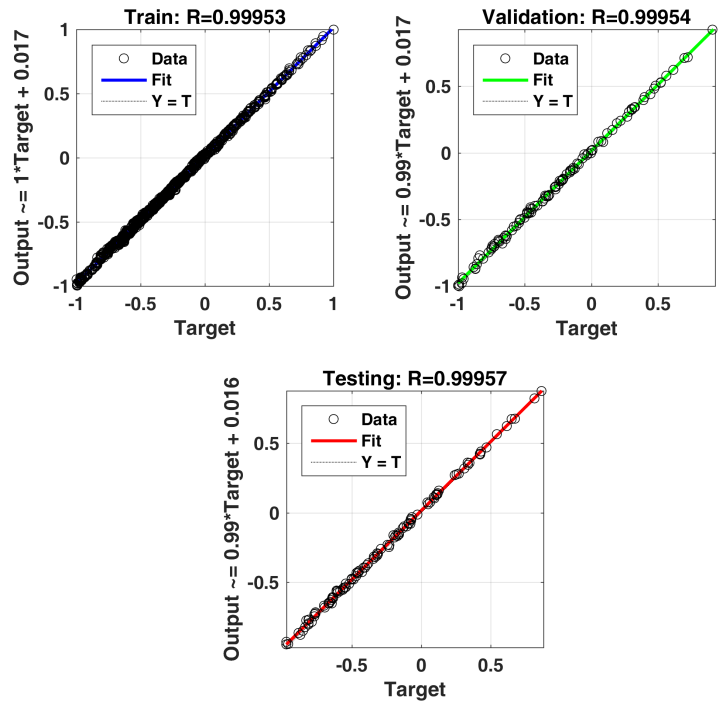

Fig. 16 Regression plots of the best neural network for ($n_{trim}$) case

As can be seen, the model developed by the nonlinear least-squares method within the neural network training framework is fitted to the training data and capable of generalizing to the test data with slightly better performance and less number of coefficients than the *Poly3344* model. However, the little superiority of the neural network in the obtained results is achieved at the higher computational cost of the iterative scheme. The parameters of the *Poly3344* model are evaluated in 0.062 seconds whereas the average total time required for 15 10-hidden neuron networks to be trained and their best model to be extracted is 10.865 seconds (this is an average value as the 15 networks are initialized from different weights and biases each time being trained). Therefore, it is notable that the linear least-squares method can almost instantaneously (i.e. in a single pass) provide a fit nearly as good as the one obtained through the nonlinear regression. This is specifically important if the regression method is to be employed in online training to account for the necessary variations of the fitted model required to be exerted in case of further alterations in the impaired aircraft nonlinear dynamics due to subsequent failures (e.g. a secondary actuator failure resulted by excessive control effort).

Although, it should be noted that this inference is based on the results obtained for the rudder failure and cannot be generalized to other types of failure such as structural damages or combinatory failures. Specifically, in the case of structural failures many of the non-dimensional derivatives which characterize the force and moment coefficients are changed from their nominal values of the unimpaired case [64]. Taking into account these derivatives along with the altitude and flight path angle as the input parameters with influence on the MFE variations leads to a considerable increase in the number of model input variables with respect to the case of rudder failure where there are only four inputs [65]. Hence, the linear least-squares method may be incapable of accurately modeling the nonlinear relationship between the MFE variations and structural failure degrees. Assuming the availability of proper system identification methods [39, 66] to estimate the aerodynamic derivatives, it is anticipating that the nonlinear regression method and specifically neural networks with their significant learning capabilities would be able to perform the aforementioned modeling in the case of structural damages. This can be studied in future researches as the specific numerical methods and wind tunnel tests required for the evaluation of the force and moment non-dimensional derivatives at various structural failure degrees are beyond the scope and available resources of this study.

Nevertheless, this research shows that instead of local prediction of flight envelope characteristics at an individual input, it is feasible to predict the key parameters characterizing MFE contractions and displacements at any a-priori unknown actuator failure degree and flight condition using a unified regression-derived model which takes the identified influential variables; altogether as inputs.

## 5. $C_{trim}$ as model output

So far, the presented results were associated with models evaluated with the number of trim points ($n_{trim}$) as the intended model response. The following tables present the specifications and corresponding results of the fitted models which estimate the centroid of the MFE ($C_{trim}$) as the output. As mentioned previously, the 2D MFEs whose centroids are the targets belong to the $(V - \dot{\psi})$ plane. Hence, each output sample is comprised of two elements: the corresponding $V$ and $\dot{\psi}$ values. The variations of each of these two output elements at different failure degrees and flight conditions are modeled with a separate polynomial. Therefore, the linear least-squares method requires two different models in case of ($C_{trim}$). On the other hand, the neural networks are capable of predicting multi-outputs. Thus, the coordinate of the MFE centroid is estimated via a single model fitted by the nonlinear regression method.

After performing the previously explained investigations on the considered polynomials and neural networks using the same divisions of the dataset as in the case of ($n_{trim}$), it was found that a *Poly3666* polynomial model and a 22-hidden neuron 2-layer feedforward network provide the best fits. Both models indicate that there are far more nonlinearities in the variations of $C_{trim}$ with the four inputs than the existing nonlinearities in the relationship between $n_{trim}$ and the same inputs. Specifically, the *Poly3666* model shows that for both of the centroid elements, even the flight path angle requires a highest degree of 6; same as the upper and lower limits of the rudder deflection angle, and a highest degree of 3 is only adequate for the altitude, whereas in the case of $n_{trim}$ a highest degree of 3 was sufficient to model the nonlinearities associated with both the altitude and the flight path angle. Also, the goodness of fit and generalization capability of the 2-layer network improves up to 22 hidden neurons. In order to be able to compare the multi-output neural network with the single-output polynomials, the same loss function used in the neural network is applied to the outputs of the two fitted polynomials as below

$$MSE = {1}/{110} \sum_{i=1}^{110} \left[ {1}/{2} \sum_{j=1}^{2} (\hat{y}_{ij} - y_{ij})^2 \right] \tag{51}$$

so that for the two *Poly3666* models a single MSE is evaluated for each of the training set and the test set. The fourth to seventh columns of Table 12 present the individual MSEs of the polynomials along with their aggregated MSEs obtained by (51).

**Table 12 Fitted models' specifications and performances ($C_{trim}$)**

| Model | Number of coefficients | Adjusted $R^2$ value | Training set MSE | Test set MSE | Total fitting time (sec) |
|---|---|---|---|---|---|
| 2-layer 22-hidden neuron Neural Network | 156 | 0.9987 | 5.8204E-5 | 6.2769E-5 | 31.246 |
| Poly3666 ($V$) | 195 | 0.9969 | 7.0313E-5 | 6.9026E-5 | 0.138 |
| Poly3666 ($\dot{\psi}$) | 195 | 0.9976 | 8.6071E-5 | 8.9118E-5 | 0.138 |
| | | | 7.8192E-5 | 7.9072E-5 | 0.276 |

**Table 13 Fitted models' error percentages for the five failure cases ($C_{trim}$)**

| Failure case | $C_{trim}$ $\dot{\psi}$ (deg/s) | $V$ (knot) | Poly3666 | | 2-layer 22-hidden neuron Neural Network | |
|---|---|---|---|---|---|---|
| (1) | -2.5664 | 104.6921 | -2.5725 | 104.0816 | -2.5985 | 104.5340 |
| | | | (0.1724%) | | (0.5496%) | |
| (2) | 2.1869 | 109.0149 | 2.2196 | 109.0132 | 2.1366 | 108.8135 |
| | | | (0.7474%) | | (1.2432%) | |
| (3) | -2.7218 | 102.6440 | -2.6931 | 102.6426 | -2.7109 | 102.2875 |
| | | | (0.5266%) | | (0.0273%) | |
| (4) | 3.7756 | 107.6299 | 3.6119 | 107.1825 | 3.7765 | 107.1785 |
| | | | (2.3757%) | | (0.2211%) | |
| (5) | -1.5837 | 123.4407 | -1.4842 | 122.9677 | -1.5241 | 122.7581 |
| | | | (2.9499%) | | (1.6059%) | |

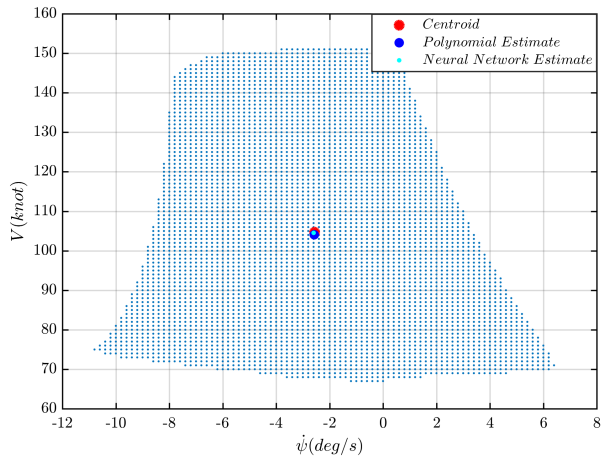

Fig. 17 Fitted models' estimates of the centroid for failure case (1)

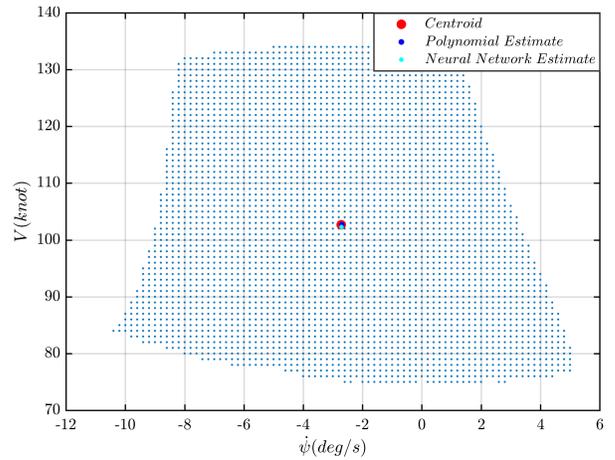

Fig. 18 Fitted models' estimates of the centroid for failure case (2)

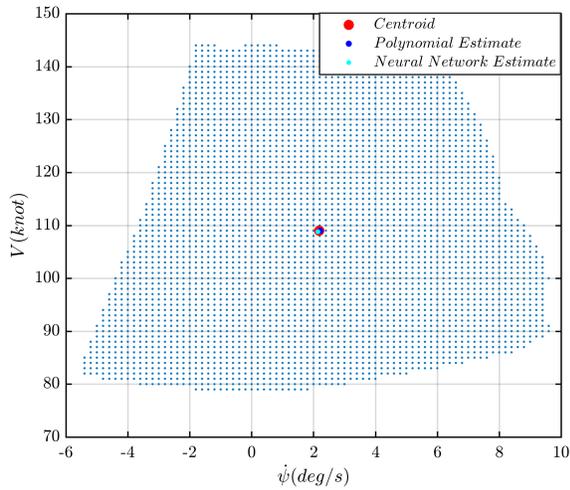
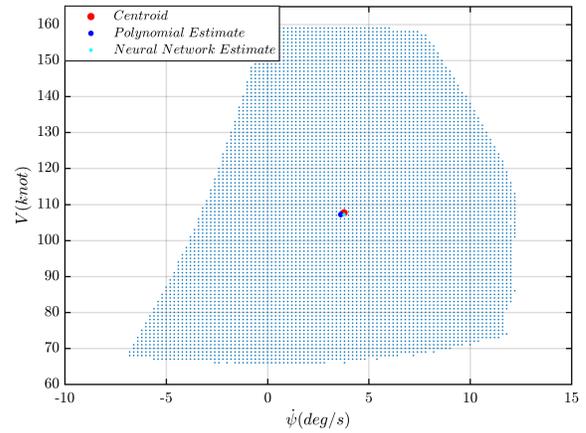

**Fig. 19 Fitted models' estimates of the centroid for failure case (3)**

**Fig. 20 Fitted models' estimates of the centroid for failure case (4)**

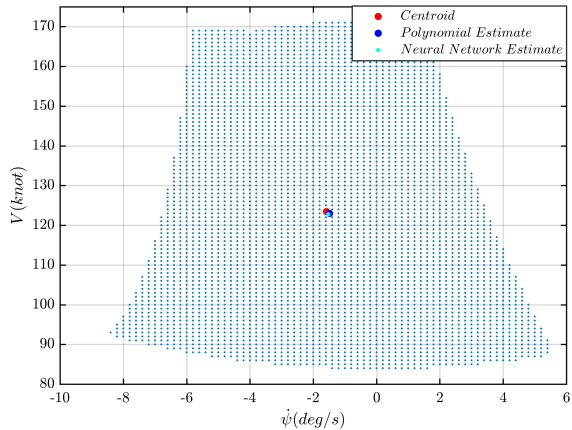

**Fig. 21 Fitted models' estimates of the centroid for failure case (5)**

The increased total degree of the fitted polynomials and hidden neurons of the fitted network have resulted in considerably higher number of coefficients. However, it is noteworthy that in order to predict the centroid of the MSE with the linear regression model, the values of a total of 390 coefficients should be estimated whilst the multi-output network has 156 weights and biases. For the failure type of this study, neural networks are up to 60% smaller than polynomials. For more complex failure types this difference could be even more, and when all failure type are modeled and considered together there would be a significant difference between the required memory of the developed neural networks and the developed polynomials. This is important as it is much more desirable to develop models with smaller sizes for aircraft onboard applications.

Despite the significant difference between the sizes of the fitted models by the linear and nonlinear least-squares techniques, the required total time to fit the models is still negligible for the polynomials (with respect to the case of ($n_{trim}$)) whereas it has almost tripled for the neural network. For better comprehension, the figures 17 to 21 depict the results for the five failure cases which indicate the acceptable generalization of both models.

## V. Sensitivity Analysis of Impaired Aircraft's Flight Envelope Variations

In order to assess the degree of effect of the designated inputs on the impaired aircraft's MFE variations, the variance-based global sensitivity analysis method has been employed which takes into account the variations of the input factors within the entire variability space and provides accurate numerical indices representing the individual and group contributions of the input variables to the MFE contraction and displacement.

Considering that the approximating model output $y$ is either the number of trim points ($n_{trim}$) or each of the two centroid elements ($V_{C_{trim}}, \dot{\psi}_{C_{trim}}$), its variance $Var(y)$ can be expressed as [67]:

$$E_{z_j}\left(Var_{z_{\sim j}}(y|z_j)\right) + Var_{z_j}\left(E_{z_{\sim j}}(y|z_j)\right) = Var(y) \tag{52}$$

In the variance-based sensitivity analysis, the contribution to the model output variance from a specific input factor is considered as a measure of sensitivity. Hence, the first term of (52) which indicates the average of the conditional variance of $y$, taken over all factors but the $j^{th}$ input factor ($z_j$) when $z_j$ is fixed, is a measure of influence of other factors but $z_j$. The smaller this term, the greater the second term of (52) and the influence of $z_j$. Therefore, the second term is the main effect of $z_j$ on $y$ with the corresponding first-order sensitivity index $\mathbb{S}_j$ presented in (53).

$$\mathbb{S}_j = Var_{z_j}\left(E_{z_{\sim j}}(y|z_j)\right)/Var(y) \tag{53}$$

Likewise, the total-effect of the input factor $z_j$ which is the aggregated contribution from the $j^{th}$ input to the output variance comprising its main (i.e. individual) effect and shared effects due to all interactions with other inputs is indicated by the total-order sensitivity index $\mathbb{S}_j^T$ and defined as [67]:

$$\mathbb{S}_j^T = E_{z_{\sim j}}\left(Var_{z_j}(y|z_{\sim j})\right)\bigg/Var(y) = 1 - Var_{z_{\sim j}}\left(E_{z_j}(y|z_{\sim j})\right)\bigg/Var(y) \tag{54}$$

First and total-order indices can be efficiently estimated via Monte-Carlo estimators within a Monte-Carlo based numerical procedure. In this research, the method proposed in [67] has been adopted to evaluate the mentioned estimators. Briefly, the procedure requires generating two $(N, \mathcal{H})$ matrices of random numbers, constructing a third $(\mathcal{H}N, \mathcal{H})$ matrix from the columns of the two $(N, \mathcal{H})$ matrices, and evaluating the model output for all $(2 + \mathcal{H})N$ samples in the three matrices. The number of base samples $N$ depends on the amount of nonlinearity and complexity of the model. Also, the number of columns $(\mathcal{H})$ of the three matrices corresponds to the number of input variables considered in the sensitivity analysis. More details on the employed Monte-Carlo estimators and the numerical procedure can be found in [67].

The variance of the model output can be decomposed as in (55) when the input factors are independent. Hence, the higher order indices corresponding to the effects of interactions between the input factors (which cannot be expressed as the sum of their main effects) can be represented for the $j^{th}$ input factor in relation to its first and total order indices as in (56).

$$Var(y) = \sum_j Var_{z_j} + \sum_j \sum_{i>j} Var_{z_{ji}} + \cdots + Var_{z_{12\ldots\mathcal{H}}} \tag{55}$$

$$\mathbb{S}_j^T = \mathbb{S}_j + \mathbb{S}_{ji} + \mathbb{S}_{jk} + \cdots + \mathbb{S}_{jik} + \cdots + \mathbb{S}_{jik\ldots h} \tag{56}$$

where $Var_{z_{ji}}$ is the joint effect of the input factors $z_j$ and $z_i$ minus their first-order effects, with the corresponding higher order index being $\mathbb{S}_{ji}$. In this study, Latin-Hypercube sampling has been used to generate the aforementioned two $(N, \mathcal{H})$ matrices. Latin-Hypercube sampling is a specific type of stratified sampling that reduces the gaps between the clusters of sampled points and thus produces a more uniform sample grid than the one generated in the random sampling strategy by the pseudo-random number generator. Figures 22 to 25 depict the first and total-order sensitivity indices computed by applying the variance-based sensitivity analysis to the developed *Poly3344* and *Poly3666* models.

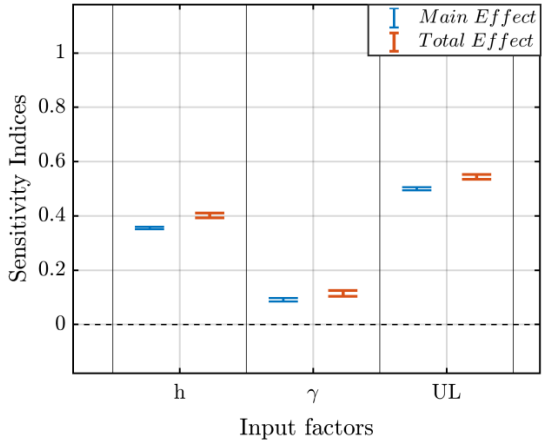

**Fig. 22 Sensitivity indices of selected inputs for the model output $n_{trim}$ (based on *Poly3344*)**

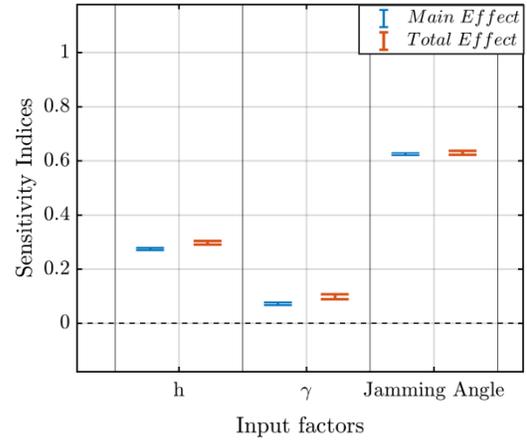

**Fig. 24 Sensitivity indices of selected inputs for the model output $n_{trim}$ (based on *Poly3344*)**

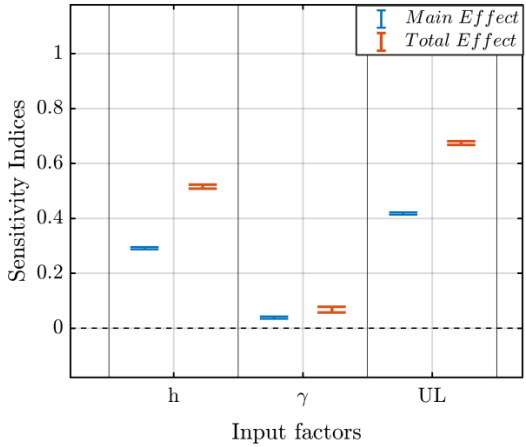

**Fig. 23 Sensitivity indices of selected inputs for the model output $\dot{\psi}_{C_{trim}}$ (based on *Poly3666*)**

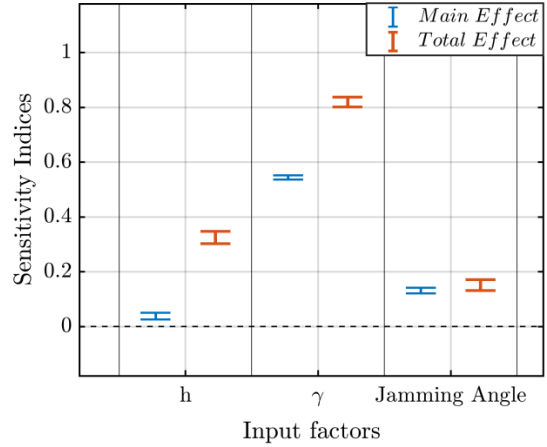

**Fig. 25 Sensitivity indices of selected inputs for the model output $V_{C_{trim}}$ (based on *Poly3666*)**

As mentioned earlier, the input variables of the developed polynomial models are $\vec{z} = [h\ \gamma\ LL\ UL]^T$. However, the deflection limits of rudder (i.e. $LL, UL$) are correlated and their values are dependent upon each other. For instance, if the lower limit of the damaged rudder's deflection angle is $-10°$, the upper limit value cannot be less than $-10°$. Therefore, the mentioned numerical procedures and corresponding Monte-Carlo estimators cannot be used unless either one of the two deflection limits is selected for the sensitivity analysis or their values are set to be equal as a jamming angle. Figures 22 and 23 indicate that the flying altitude has larger effect on the reduction of the remaining feasible trim points than the flight path angle. That is because the altitude variation affects both the aircraft's available thrust and stall speed (due to change in the air density), causing envelope contraction from top and bottom, whereas changing the flight path angle alters the amount of the required thrust and subsequently the maximum speed of the aircraft, resulting in the envelope contraction only from the top. According to figures 22 and 23, an initially unimpaired aircraft suffers larger envelope contractions due to the variation of the upper limit from its nominal value of 30° (whilst the lower limit is fixed at the nominal value of −30°) than the contractions caused by changing the two affecting flight condition parameters (i.e. $h$ and $\gamma$). Similarly, an already impaired aircraft with restricted rudder failure or jammed rudder failure experiences larger envelope contractions due to a secondary rudder failure such as the variation of the upper limit from its initial restricted value or due to the variation in the jamming angle. However, the jamming angle is more influential than the upper limit. In addition to the first and total order indices, numerical values of the higher order indices are presented in Table 14.

**Table 14 Numerical values of the calculated sensitivity indices**

| Fig. 22 | | | | Fig. 23 | | |
|---|---|---|---|---|---|---|
| $\mathbb{S}_h$ | 0.3565 | $\mathbb{S}_h^T$ | 0.4021 | $\mathbb{S}_h$ | 0.2751 | $\mathbb{S}_h^T$ | 0.2982 |
| $\mathbb{S}_\gamma$ | 0.0912 | $\mathbb{S}_\gamma^T$ | 0.1141 | $\mathbb{S}_\gamma$ | 0.0726 | $\mathbb{S}_\gamma^T$ | 0.0995 |
| $\mathbb{S}_{UL}$ | 0.5005 | $\mathbb{S}_{UL}^T$ | 0.5443 | $\mathbb{S}_{Jammed}$ | 0.6255 | $\mathbb{S}_{Jammed}^T$ | 0.6316 |
| $\mathbb{S}_{h\gamma}$ | 0.0125 | $\mathbb{S}_{UL\gamma}$ | 0.0104 | $\mathbb{S}_{h\gamma}$ | 0.0221 | $\mathbb{S}_{Jammed\ \gamma}$ | 0.0048 |
| $\mathbb{S}_{hUL}$ | 0.0331 | | | $\mathbb{S}_{h\ Jammed}$ | 0.0011 | | |
| Fig. 24 | | | | Fig. 25 | | | |
| $\mathbb{S}_h$ | 0.2913 | $\mathbb{S}_h^T$ | 0.5169 | $\mathbb{S}_h$ | 0.0388 | $\mathbb{S}_h^T$ | 0.3267 |
| $\mathbb{S}_\gamma$ | 0.0394 | $\mathbb{S}_\gamma^T$ | 0.0683 | $\mathbb{S}_\gamma$ | 0.5303 | $\mathbb{S}_\gamma^T$ | 0.8209 |
| $\mathbb{S}_{UL}$ | 0.4182 | $\mathbb{S}_{UL}^T$ | 0.6737 | $\mathbb{S}_{Jammed}$ | 0.1307 | $\mathbb{S}_{Jammed}^T$ | 0.1509 |
| $\mathbb{S}_{h\gamma}$ | $\cong 0$ | $\mathbb{S}_{UL\gamma}$ | 0.0290 | $\mathbb{S}_{h\gamma}$ | 0.2786 | $\mathbb{S}_{Jammed\ \gamma}$ | 0.0120 |
| $\mathbb{S}_{hUL}$ | 0.2256 | | | $\mathbb{S}_{h\ Jammed}$ | 0.0081 | | |

Such values demonstrate the existence of interactions between the input variables of the case studies of figures 22 and 23. For instance, at relatively high flying altitudes (about 30000 ft for the GTM; based on the evaluated MFEs of the database) where the thrust-available and thrust-required curves are close, a specific increase in the flight path angle results in larger contraction of the flight envelope than that caused by the same change in the flight path angle at lower altitudes. Based on the figures 24 and 25, and the corresponding numerical values in Table 14, variation of the flight path angle has little impact on the horizontal displacement of the MFE and the largest effect on its vertical displacement. This is expected as varying the flight envelope causes envelope shrinkage from the top which relocates the centroid almost vertically in the $(V - \dot{\psi})$ plane. Conversely, changing the altitude results in non-negligible horizontal displacement of the MFE's centroid but does not affect its vertical location. Also, variation of the rudder's deflection upper limit significantly changes the $\dot{\psi}$ value of the centroid as its reduction causes feasible trim points elimination from the side of the MFE. Similarly, a change in the jamming angle shifts the MFE more horizontally rather than vertically. It should be noted that the figures 24 and 25 present two sensitivity analysis instances performed on the models estimating $\dot{\psi}_{C_{trim}}$ and $V_{C_{trim}}$ in which the upper limit and the jamming angle have been selected as the third input variable, respectively. However, other third variables could also be considered, such as the lower limit or the jamming angle for the analysis of the model output $\dot{\psi}_{C_{trim}}$.

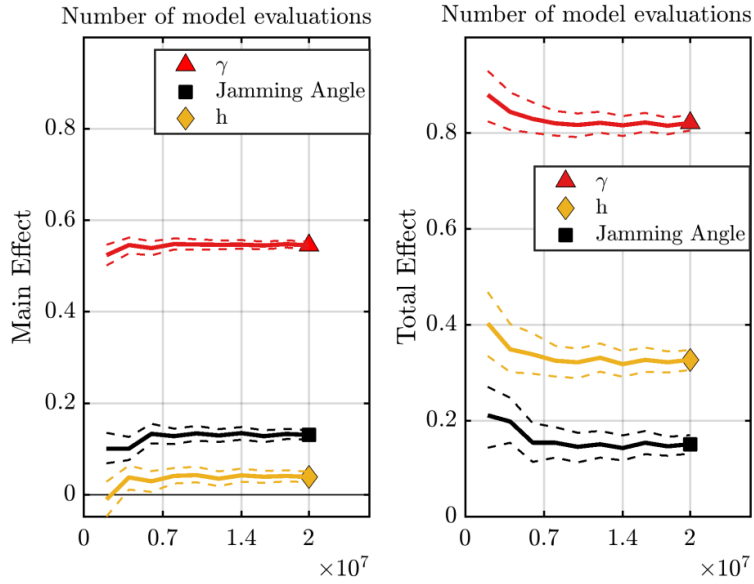

**Fig. 26 Convergence analysis of first and total-order indices for the model output $V_{C_{trim}}$ (based on *Poly3666*)**

As mentioned earlier, the employed numerical procedure requires $5N$ model evaluations for each sensitivity analysis consisting three input variables. The indices' narrow confidence intervals shown in figures 22 to 25 were obtained by increasing the number of base samples ($N$) to 500,000 for the analyses of figures 22 to 24 and to 4,000,000 for the analysis of Fig. 25. Therefore, a total of 22,500,000 model evaluations were conducted for the variance-based analyses of figures 22 to 25. Due to well-structured regression models which directly relate the considered envelope characteristics to the intended input variables, such a large number of model evaluations was computationally affordable. To be specific, approximately 600,000 model evaluations were executed per minute on a standard computer with 2.20 GHz Intel® Core i7-8750H processor and 32 GB RAM, under Windows 10 operating system, and using MATLAB® version 9.3 (R2017b). The robustness of the estimated indices was evaluated by deriving the corresponding confidence intervals via percentile bootstrap re-sampling [68]. For instance, Fig. 26 presents the convergence analysis results for the case study of Fig. 25.

Presented regression-based global sensitivity analysis approach enables assessing the degree of effect of different contributing parameters to the variations of the impaired aircraft's maneuvering flight envelope. Such evaluation provides prior knowledge of the most effective parameters limiting the impaired aircraft's maneuverability, which can be used as an advisory in specifying post-failure path planning strategies. This is specifically useful in the case of combinatory failures such as an impaired aircraft with damaged longitudinal and lateral control surfaces where identifying the most limiting input factor is not as intuitive as the case studies presented above. In that case, comparing the sensitivity indices of the defective control surfaces' deflection limits determines the surface whose further restriction confines the maneuvering flight envelope more dominantly. Considering that, the post-failure path planning objective could be chosen such that minimum control effort is imposed on the most influential damaged control surface. For instance, a level turn maneuver in the horizontal plane could be selected over a climbing-descending maneuver in the vertical plane to avoid the ahead-terrain when flight envelope contraction is more susceptible to the damaged elevator rather than the damaged rudder. This ensures that the flight envelope would not degrade too much in the case of a secondary failure of the damaged control surface.

## VI.     Conclusion

In this paper, various nonlinear models are developed by the linear and nonlinear least-squares methods with the aim of predicting the number of remaining feasible trim points and the centroid location of the impaired NASA GTM's shrunk and displaced maneuvering flight envelope at any a-priori unknown failure degree. These two model outputs which characterize the maneuvering flight envelope variation of an impaired aircraft are dependent on the flight altitude, flight path angle, and specific parameters determined by the failure type. These failure-related parameters are the lower and upper limits of the defective control surface's deflection angle in the considered failure cases of this study. Whilst generally there is no explicit relationship between the aforementioned influential input parameters and the two intended outputs, the developed nonlinear models of this study provide analytical functions capable of estimating the outputs at any intended failure degree with high precision. Specifically, polynomials of different degrees (as the linear regression models) and 2-layer feedforward neural networks with different number of hidden neurons (as the nonlinear regression models) are built and investigated to find the best fitting polynomial and neural network. According to the obtained results, a 4$^{\text{th}}$ degree polynomial and a 10-hidden neuron network are able to accurately predict the remaining number of trim points whereas the centroid location of the impaired flight envelope can be estimated with a 6$^{th}$ degree polynomial and a 22-hidden neuron network. More complex models are required in case of the centroid estimation due to higher nonlinearities between the inputs and the output. By comparing the developed models based on their mean squared errors on a test dataset it was found that the best neural networks have slightly better performance and smaller size (i.e. lower number of coefficients) than the best polynomials. However, unlike the neural networks, the parameters of the polynomials can be evaluated almost instantaneously, which is not crucial in case the models are developed offline, but could be advantageous if the developed models' parameters are supposed to be further modified online during the flight. In either case, the obtained results demonstrate the feasibility of predicting the key parameters characterizing flight envelope contraction and displacement at any a-priori unknown actuator failure degree and flight condition using a unified regression-derived model which takes the identified influential variables; altogether as inputs. Estimating the horizontal and vertical extent of the flight envelope around its centroid as a 3D shape using the fast linear least-squares method, and investigating the possibility of developing similar models for other failure types such as structural damages are potential topics for future researches.

Also, it is shown in this study that the regression equations of the developed polynomials can be used as fast and accurate emulators for millions of model evaluations required for estimating the sensitivity indices of the input variables through a global sensitivity analysis approach. The results of four performed sensitivity analyses on the MFE's number of trim point and centroid elements are presented and discussed. The most effective input variables are identified, and it is explained how these results can be interpreted for the post-failure path planning.

## VII.    Appendix A

This section presents the results of the *poly4444* model and compares them with the other constructed models. According to the obtained results presented in the first three rows of the Table 15, it seems that higher degree models better fit the data as the *adjusted $R^2$* increases, and both the MSEs on the training data and on the test data decrease. Specifically, the values of the MSEs suggest that the *poly4444* model not only fits the training data better than the other two models but it also generalizes very well to the new unseen data. However, the evaluated models' error percentages on the five failure cases presented in Table 16 contradict this conclusion. In fact, the error percentages of the first two failure cases ((1) and (2)) – which belong to the test set – decrease by increasing the polynomial's degree, however, for the rest three cases – which as mentioned earlier have been selected from the mid-range values of the input variables – the *poly4444* model's outcomes are largely erroneous. Hence, it can be inferred that the values of the *adjusted* $R^2$ and the MSEs on the training and test sets cannot guarantee the generalization of the polynomial model and further random samples' check and visualization of the fitted model are required for diagnosing and improving the model.

Since the fitted model is five dimensional, it cannot be depicted unless two of the four input variables are fixed. Figure 27 presents the fitted *poly4444* model in which the *LL* has been set equal to the *UL* and $\gamma = 2°$. In other words, the figure depicts the variation of the number of trim points $n_{trim}$ with various altitudes and different jamming angles at $\gamma = 2°$. Even though the model has fitted the training data (red points) very well, for the test samples with altitude values in between the altitudes of the training data the model predictions are quite divergent. This is due to the oscillatory interpolation property which is one of the disadvantages of the polynomials [69].

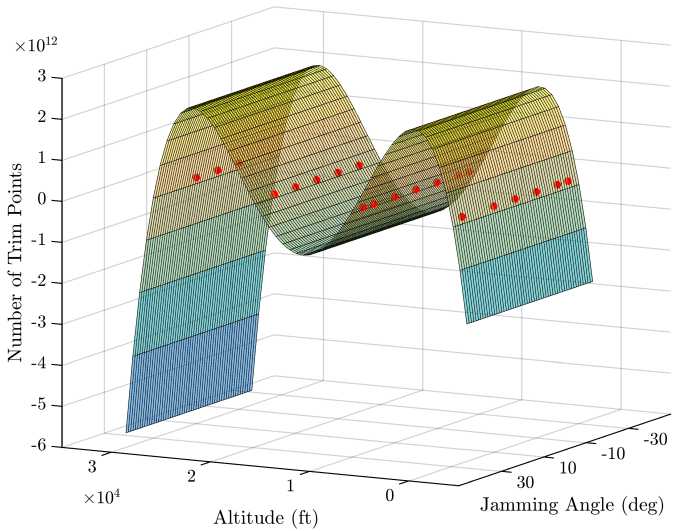

**Fig. 27 3D visualization of the 4$^{th}$ degree polynomial model at $\gamma = 2°$ (Poly4444)**

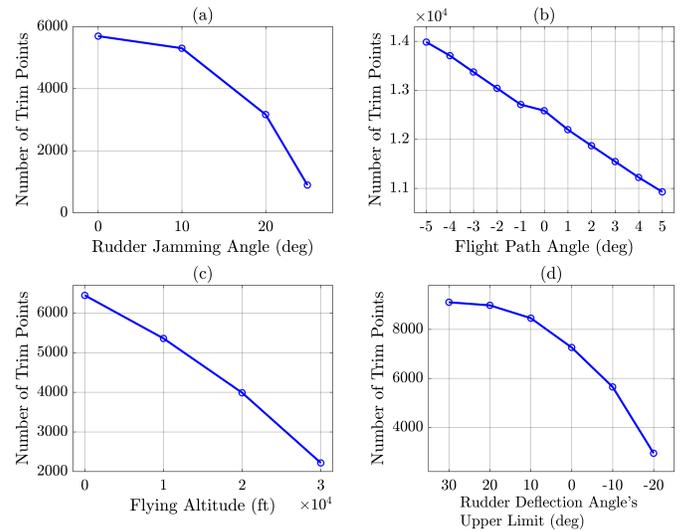

**Fig. 28 Variations of observed $n_{trim}$ with changes in the input variables (training data)**

This shows that the fitted *poly4444* model includes certain variable degrees that are beyond the amount of nonlinearities in the corresponding variables. Those variables must be identified and their degrees must be lowered such that the unnecessary polynomial terms which do not account for the actual variations and nonlinearities are omitted. As can be seen in Fig. 28, the variations of $n_{trim}$ with altitude and flight path angle are almost linear, whereas it varies nonlinearly with the jamming angle (*LL=UL*) and the *UL* itself. Therefore, the maximum degrees of $h$ and $\gamma$ in the fitting model terms must be less than those of the *UL* and *LL*. As mentioned earlier, in the fitted *poly4444* model all the input variables have a maximum degree of 4. After investigating alternative models with lower maximum-degrees of $h$ and $\gamma$, it was found that the best model is a *poly3344* wherein the maximum degrees of $h$ and $\gamma$ are 3. In fact, the *poly3344* model has only two fewer terms (model coefficients) than the *poly4444* model as the $a_{4000}h^4\gamma^0 LL^0 UL^0$ and $a_{0400}h^0\gamma^4 LL^0 UL^0$ terms are omitted. It should be noted that there are no terms in the *poly4444* model consisting of non-zero degrees of the *UL* and *LL* along with 4$^{th}$ degree of $h$ or $\gamma$, as such terms' degrees sum to more than the total degree of the polynomial (i.e. 4 in the case of *poly4444* model) which is the maximum of the input variables' highest degrees.

The *poly3344* model not only has lower training and test MSEs than the presented 2$^{nd}$ and 3$^{rd}$ degree polynomials but also fits the aforementioned five failure cases better than the *poly4444* model. Tables 15 and 16 present the *poly3344* model specifications and results in comparison with the other fitted models.

**Table 15 *Poly3344* model's specifications and MSEs against other models**

| Model | Number of coefficients | Adjusted $R^2$ value | Degrees of freedom (DOF) | Training set MSE | Test set MSE |
|---|---|---|---|---|---|
| Poly2222 | 15 | 0.9884 | 976 | 5.0251E-4 | 5.5858E-4 |
| Poly3333 | 35 | 0.9948 | 956 | 2.1851E-4 | 2.1058E-4 |
| Poly4444 | 70 | 0.9979 | 921 | 8.4400E-5 | 7.5933E-5 |
| Poly3344 | 68 | 0.9976 | 923 | 9.7601E-5 | 9.5569E-5 |

**Table 16 *Poly3344* model's error percentages for the five failure cases against other models**

| Failure case | $n_{trim}$ | Poly2222 | Poly3333 | Poly4444 | Poly3344 |
|---|---|---|---|---|---|
| (1) | 4898 | 4483 | 4756 | 4825 | 4852 |
| | | (8.4645%) | (2.8975%) | (1.4884%) | (0.9436%) |
| (2) | 3759 | 3378 | 3533 | 3825 | 3786 |
| | | (10.1229%) | (6.0117%) | (1.7542%) | (0.7159%) |
| (3) | 3508 | 2847 | 3109 | -1.1493e12 | 3285 |
| | | (18.8338%) | (11.3663%) | (3.2764e10%) | (6.3497%) |
| (4) | 6538 | 6341 | 6445 | 9.0249e11 | 6605 |
| | | (3.0140%) | (1.4187%) | (1.3804e10%) | (1.0205%) |
| (5) | 4348 | 4063 | 4157 | 1.3427e12 | 4208 |
| | | (6.5507%) | (4.3960%) | (3.0880e10%) | (3.2164%) |

## Funding Sources

This research did not receive any specific grant from funding agencies in the public, commercial, or not-for-profit sectors.

# References

[1] "Statistical Summary of Commercial Jet Airplane Accidents, Worldwide Operations | 1959 – 2014," Aviation Safety, Boeing Commercial Airplanes, Seattle, WA, October 2018.

[2] "Global Fatal Accident Review 2002 - 2011," TSO (The Stationery Office) on behalf of the UK Civil Aviation Authority, Norwich, UK, 2013 [Online]. Available: http://publicapps.caa.co.uk/cap1036. [retrieved 9 February 2018].

[3] Gill, S. J., Lowenberg, M. H., Neild, S. A., Krauskopf, B., Puyou, G., and Coetzee, E., "Upset Dynamics of an Airliner Model: A Nonlinear Bifurcation Analysis," *Journal of Aircraft*, vol. 50, no. 6, Nov. 2013, pp. 1832–1842.
doi: 10.2514/1.C032221

[4] Norouzi, R., Kosari, A., and Sabour, M. H., "Investigating the Generalization Capability and Performance of Neural Networks and Neuro-Fuzzy Systems for Nonlinear Dynamics Modeling of Impaired Aircraft," *IEEE Access*, vol. 7, Feb. 2019, pp. 21067–21093
doi: 10.1109/ACCESS.2019.2897487

[5] Yi, G., and Atkins, E., "Trim State Discovery for an Adaptive Flight Planner," *48th AIAA Aerospace Sciences Meeting Including the New Horizons Forum and Aerospace Exposition*, AIAA Paper 2010-416, Jan. 2010.
doi: 10.2514/6.2010-416

[6] Tang, Y., Atkins, E., and Sanner, R., "Emergency Flight Planning for a Generalized Transport Aircraft with Left Wing Damage," *AIAA Guidance, Navigation and Control Conference and Exhibit*, AIAA Paper 2007-6873, Aug. 2007.
doi: 10.2514/6.2007-6873

[7] Frink N. T., Murphy, P. C., Atkins, H. L., Viken, S. A., Petrilli, J. L., Gopalarathnam, A., Paul, R. C., "Computational Aerodynamic Modeling Tools for Aircraft Loss of Control," *Journal of Guidance, Control, and Dynamics*, vol. 40, no. 4, pp. 789–803, Apr. 2017
doi: 10.2514/1.G001736

[8] Wilborn, J., and Foster, J., "Defining Commercial Transport Loss-of-Control: A Quantitative Approach," *AIAA Atmospheric Flight Mechanics Conference and Exhibit*, AIAA Paper 2004-4811, Aug. 2004.
doi: 10.2514/6.2004-4811

[9] Vannelli, A., and Vidyasagar, M., "Maximal Lyapunov functions and domains of attraction for autonomous nonlinear systems," *Automatica*, vol. 21, no. 1, Jan. 1985, pp. 69–80.
doi: 10.1016/0005-1098(85)90099-8

[10] Chiang, H.-D., Hirsch, M. W., and Wu, F. F., "Stability regions of nonlinear autonomous dynamical systems," *IEEE Transactions on Automatic Control*, vol. 33, no. 1, Jan. 1988, pp. 16–27.
doi: 10.1109/9.357

[11] Chesi, G., Garulli, A., Tesi, A., and Vicino, A., "LMI-based computation of optimal quadratic Lyapunov functions for odd polynomial systems," *International Journal of Robust and Nonlinear Control*, vol. 15, no. 1, 2004, pp. 35–49.

doi: 10.1002/rnc.967

[12] Tibken, B., "Estimation of the domain of attraction for polynomial systems via LMIs," *Proceedings of the 39th IEEE Conference on Decision and Control* (Cat. No.00CH37187), IEEE, Dec. 2000.

doi: 10.1109/CDC.2000.912314

[13] Amato, F., Cosentino, C., and Merola, A., "On the region of attraction of nonlinear quadratic systems," *Automatica*, vol. 43, no. 12, Dec. 2007, pp. 2119–2123.

doi: 10.1016/j.automatica.2007.03.022

[14] Pandita, R., Chakraborty, A., Seiler, P., and Balas, G., "Reachability and Region of Attraction Analysis Applied to GTM Dynamic Flight Envelope Assessment," *AIAA Guidance, Navigation, and Control Conference*, AIAA Paper 2009-6258, Aug. 2009.

doi: 10.2514/6.2009-6258

[15] Zheng, W., Li, Y., Zhang, D., Zhou, C., and Wu, P., "Envelope protection for aircraft encountering upset condition based on dynamic envelope enlargement," *Chinese Journal of Aeronautics*, vol. 31, no. 7, pp. 1461–1469, Jul. 2018

doi: 10.1016/j.cja.2018.05.006

[16] Yuan, G., and Li, Y., "Determination of the flight dynamic envelope via stable manifold," *Measurement and Control*, vol. 52, no. 3–4, pp. 244–251, Feb. 2019

doi: 10.1177/0020294019830115

[17] Lombaerts, T., Schuet, S., Wheeler, K., Acosta, D., and Kaneshige, J., "Robust Maneuvering Envelope Estimation Based on Reachability Analysis in an Optimal Control Formulation," *Conference on Control and Fault-Tolerant Systems* (SysTol), IEEE Publ., Piscataway, NJ, Oct. 2013, pp. 318–323.

doi: 10.1109/SysTol.2013.6693856

[18] Harno, H. G., Kim, Y., "Safe Flight Envelope Estimation for Rotorcraft: A Reachability Approach," *18th International Conference on Control, Automation, Systems* (ICCAS), IEEE Publ., Daegwallyeong, South Korea, Dec. 2018.

[19] Tang, L., Roemer, M., Ge, J., Crassidis, A., Prasad, J., and Belcastro, C., "Methodologies for Adaptive Flight Envelope Estimation and Protection," *AIAA Guidance, Navigation, and Control Conference*, AIAA Paper 2009-6260, Aug. 2009.

[20] Oort, E. V., Chu, P., and Mulder, J. A., "Maneuvering Envelope Determination Through Reachability Analysis," *Advances in Aerospace Guidance, Navigation and Control*, edited by Holzapfel, F., and Theil, S., Springer–Verlag, Heidelberg, 2011, pp. 91–102.

doi: 10.1007/978-3-642-19817-5_8

[21] Kampen, E., Chu, Q. P., Mulder, J. A., and Emden, M. H., "Nonlinear aircraft trim using internal analysis," *AIAA Guidance, Navigation and Control Conference and Exhibit*, AIAA Paper 2007-6766, Aug. 2007.

doi: 10.2514/6.2007-6766

[22] Goman, M. G., Khramtsovsky, A. V., and Kolesnikov, E. N., "Evaluation of Aircraft Performance and Maneuverability by Computation of Attainable Equilibrium Sets," *Journal of Guidance, Control, and Dynamics*, vol. 31, no. 2, Mar. 2008, pp. 329–339.

doi: 10.2514/1.29336

[23] Strube, M. J., Sanner, R., and Atkins, E., "Dynamic Flight Guidance Recalibration after Actuator Failure," *AIAA 1st Intelligent Systems Technical Conference*, AIAA Paper 2004-6255, Sep. 2004.

doi: 10.2514/6.2004-6255

[24] Strube, M. J., "Post-failure trajectory planning from feasible trim state sequences," M.S. thesis, Dept. Aerospace Eng. Univ. Maryland, College Park, MD, 2005.

[25] Choi, H. J., Atkins, E., and Yi, G., "Flight Envelope Discovery for Damage Resilience with Application to an F-16," *AIAA Infotech@Aerospace*, AIAA Paper 2010-3353, April 2010.

doi: 10.2514/6.2010-3353

[26] Asadi, D., Sabzehparvar, M., and Talebi, H. A., "Damaged airplane flight envelope and stability evaluation," *Aircraft Engineering and Aerospace Technology*, vol. 85, no. 3, May 2013, pp. 186–198.

doi: 10.1108/00022661311313623

[27] Asadi, D., Sabzehparvar, M., Atkins, E., and Talebi, H. A., "Damaged Airplane Trajectory Planning Based on Flight Envelope and Motion Primitives," *Journal of Aircraft*, vol. 51, no. 6, Nov. 2014, pp. 1740–1757.

doi: 10.2514/1.C032422

[28] Yi, G., Zhong, J., Atkins, E., and Wang, C., "Trim State Discovery with Physical Constraints," *Journal of Aircraft*, vol. 52, no. 1, Jan. 2015, pp. 90–106.

doi: 10.2514/1.C032619

[29] Frazzoli, E., Dahleh, M. A., and Feron, E., "Real-Time Motion Planning for Agile Autonomous Vehicles," *Journal of Guidance, Control, and Dynamics*, vol. 25, no. 1, Jan. 2002, pp. 116–129.

doi: 10.2514/2.4856

[30] Kwatny, H. G., and Allen, R. C., "Safe Set Maneuverability of Impaired Aircraft," *Guidance, Navigation, and Control and Co-Located Conferences*, AIAA Paper 2012-4405, Aug. 2012.

doi: 10.2514/6.2012-4405

[31] Yinan, L., Lingyu, Y., and Gongzhang, S., "Steady maneuver envelope evaluation for aircraft with control surface failures," *IEEE Aerospace Conference*, IEEE, March 2012.

doi: 10.1109/AERO.2012.6187317

[32] Kitsios, I., Lygeros, J., "Launch-Pad Abort Flight Envelope Computation for a Personnel Launch Vehicle Using Reachability," *AIAA Guidance, Navigation, and Control Conference and Exhibit*, AIAA Paper 2005-6150, 2005.

doi: 10.2514/6.2005-6150

[33] Nabi, H. N., Lombaerts, T., Zhang, Y., van Kampen, E., Chu, Q. P., de Visser, C. C., "Effects of Structural Failure on the Safe Flight Envelope of Aircraft," *Journal of Guidance, Control, and Dynamics*, vol. 41, no. 6, Jun. 2018, pp. 1257–1275.

doi: 10.2514/1.G003184

[34] Di Donato, P. F. A., Balachandran, S., McDonough, K., Atkins, E., Kolmanovsky, I., "Envelope-Aware Flight Management for Loss of Control Prevention Given Rudder Jam," *2010 Journal of Guidance, Control, and Dynamics*, vol. 40, no. 4, Apr. 2017, pp. 1027–1041.

doi: 10.2514/1.G000252

[35] Zhang, Y., de Visser, C. C., and Chu, Q. P., "Online Safe Flight Envelope Prediction for Damaged Aircraft: A Database-Driven Approach," *AIAA Modeling and Simulation Technologies Conference*, AIAA SciTech, AIAA Paper 2016-1189, 2016.

doi: 10.2514/6.2016-1189

[36] Zhang, Y., de Visser, C. C., and Chu, Q. P., "Aircraft Damage Identification and Classification for Database-Driven Online Safe Flight Envelope Prediction," *AIAA Atmospheric Flight Mechanics Conference*, AIAA SciTech, AIAA Paper 2017-1863, 2017.

doi: 10.2514/6.2017-1863

[37] Zhang, Y., de Visser, C. C., Chu, Q. P., "Database Building and Interpolation for a Safe Flight Envelope Prediction System," *2018 AIAA Information Systems-AIAA Infotech @ Aerospace*, Jan. 2018.

doi: 10.2514/6.2018-1635

[38] Lombaerts, T., Schuet, S., Acosta, D., Kaneshige, J., Shish, K., Martin, L., "Piloted Simulator Evaluation of Safe Flight Envelope Display Indicators for Loss of Control Avoidance," *Journal of Guidance, Control, and Dynamics*, vol. 40, no. 4, Apr. 2017, pp. 948–963.

doi: 10.2514/1.G001740

[39] Schuet, S., Lombaerts, T., Acosta, D., Kaneshige, J., Wheeler, K., Shish, K., "Autonomous Flight Envelope Estimation for Loss-of-Control Prevention," *Journal of Guidance, Control, and Dynamics,* vol. 40, no. 4, Apr. 2017, pp. 847–862.

doi: 10.2514/1.G001729

[40] Menon, P.K., Sengupta, P., Vaddi, S., Yang, B.-J., Kwan, J., "Impaired Aircraft Performance Envelope Estimation," *Journal of Aircraft*, vol. 50, no. 2, Mar. 2013, pp. 410–424.

doi: 10.2514/1.C031847

[41] McDonough, K., Kolmanovsky, I., "Fast Computable Recoverable Sets and Their Use for Aircraft Loss-of-Control Handling," *Journal of Guidance, Control, and Dynamics*, vol. 40, no. 4, Apr. 2017, pp. 934–947.

doi: 10.2514/1.G001747

[42] Venkataraman, R., Bauer, P., Seiler, P., Vanek, B., "Comparison of fault detection and isolation methods for a small unmanned aircraft," *Control Engineering Practice*, vol. 84, Mar. 2019, pp. 365–376.

doi: 10.1016/j.conengprac.2018.12.002

[43] Foster, J., Cunningham, K., Fremaux, C., Shah, G., Stewart, E., Rivers, R., Wilborn, J., and Gato, W., "Dynamics Modeling and Simulation of Large Transport Airplanes in Upset Conditions," *AIAA Guidance, Navigation, and Control Conference and Exhibit*, AIAA Paper 2005-5933, Aug. 2005.

doi: 10.2514/6.2005-5933

[44] "GTM_DesignSim," v1308, NASA, URL: https://github.com/nasa/GTM_DesignSim [retrieved 2 May 2018].

[45] Jordan, T., Langford, W., and Hill, J., "Airborne Subscale Transport Aircraft Research Testbed - Aircraft Model Development," *AIAA Guidance, Navigation, and Control Conference and Exhibit*, AIAA Paper 2005-6432, Aug. 2005.

doi: 10.2514/6.2005-6432

[46] Norouzi, R., Kosari, A., Sabour, M. H., "Data for: Maneuvering Flight Envelope Evaluation and Analysis of Generic Transport Model with Control Surfaces Failures," *Mendeley Data*, Feb. 2018.

doi: 10.17632/k4ntmx43x5.1

[47] Stevens, B., Lewis, F., and Johnson, E., *Aircraft control and simulation*, 3rd ed. Hoboken: Wiley, 2016, Chap. 3.

[48] Khalil, H., *Nonlinear systems*, 3rd ed., Upper Saddle River, N.J., USA: Prentice Hall, 2002.

[49] Sadraey, M., *Aircraft performance analysis*, Saarbrücken, Germany: VDM Verlag Dr. Müller, 2009, ch. 9, sec. 6.

[50] Edwards C, Lombaerts T, and Smaili H, *Fault tolerant flight control*. Berlin: Springer, 2010, Chap. 14, Sec. 3.6.

[51] Boyd, S., Vandenberghe, L., *Introduction to Applied Linear Algebra*, 1st ed. Cambridge University Press, Cambridge, UK, 2018.

doi: 10.1017/9781108583664.

[52] Trefethen, L. N., Bau, D., *Numerical Linear Algebra*, SIAM, Philadelphia, PA, 1997.

[53] Cunis, T., Burlion, L., Condomines, J.-P., "Piecewise Polynomial Modeling for Control and Analysis of Aircraft Dynamics Beyond Stall," *Journal of Guidance, Control, and Dynamics*, vol. 42, no. 4, Apr. 2019, pp. 949–957.

doi: 10.2514/1.G003618

[54] Chakraborty, A., Seiler, P., Balas, G.J., "Fast Nonlinear region of attraction analysis for flight control verification and validation," *Control Engineering Practice*, vol. 19, no. 4, Apr. 2011, pp. 335–345.

doi: 10.1016/j.conengprac.2010.12.001.

[55] Cunis, T., Burlion, L., Condomines, J.-P., "Piece-wise Identification and Analysis of the Aerodynamic Coefficients, Trim Conditions, and Safe Sets of the Generic Transport Model," *2018 AIAA Guidance, Navigation, and Control Conference*, AIAA Paper 2018-1114, Jan. 2018.

doi: 10.2514/6.2018-1114

[56] Byrd, R. H., Schnabel, R. B., Shultz, G. A., "Approximate Solution of the Trust Region Problem by Minimization over Two-Dimensional Subspaces," *Mathematical Programming*, Vol. 40, 1988, pp 247–263.

doi: 10.1007/BF01580735

[57] Kaasschieter, E.F., "Preconditioned conjugate gradients for solving singular systems," *Journal of Computational and Applied Mathematics*, Vol. 24, no. 1-2, Nov. 1988, pp 265–275.

doi: 10.1016/0377-0427(88)90358-5

[58] Hagan, M. T., Demuth, H. B., Beale, M. H., De Jesus, O., *Neural Network Design*, 2nd ed., USA, 2014.

[59] Marquardt, D. W., "An Algorithm for Least-Squares Estimation of Nonlinear Parameters," *Journal of the Society for Industrial and Applied Mathematics*, Vol. 11, 1963, pp 431–441.

doi: 10.1137/0111030

[60] Haykin, S., *Neural Networks, A Comprehensive Foundation*, 2nd ed., Pearson Education, India, 2005.

[61] Sugumaran, V., *Recent Advances in Intelligent Technologies and Information Systems*, IGI Global, India, 2015, Chap. 11.

doi: 10.4018/978-1-4666-6639-9

[62] Hornik, K., Stinchcombe, M., White, H., "Multilayer feedforward networks are universal approximators," *Neural Networks*, Vol. 2, 1989, pp 359–366.

doi: 10.1016/0893-6080(89)90020-8

[63] Nguyen, D., Widrow, B., "Improving the learning speed of 2-layer neural networks by choosing initial values of the adaptive weights," *1990 IJCNN International Joint Conference on Neural Networks*, Jun. 1990.

doi: 10.1109/IJCNN.1990.137819

[64] Ouellette, J., Raghavan, B., Patil, M., Kapania, R., "Flight Dynamics and Structural Load Distribution for a Damaged Aircraft," *AIAA Atmospheric Flight Mechanics Conference*, AIAA Paper 2009-6153, Aug. 2009.

doi: 10.2514/6.2009-6153

[65] Norouzi, R., Kosari, A., Sabour, M.H., "Real Time Estimation of Impaired Aircraft Flight Envelope using Feedforward Neural Networks," *Aerospace Science and Technology*, May 2019.

doi: 10.1016/j.ast.2019.04.048

[66] Moszczynski, G.J., Leung, J.M., Grant, P.R., "Robust Aerodynamic Model Identification: A New Method for Aircraft System Identification In the Presence of General Dynamic Model Deficiencies," *AIAA Scitech 2019 Forum*, Jan. 2019.

doi: 10.2514/6.2019-0433

[67] Saltelli, A., Ratto, M., Andres, T., Campolongo, F., Cariboni, J., Gatelli, D., Saisana, M., and Tarantola, S., *Global Sensitivity Analysis, The Primer*: Wiley, 2016.

[68] Pianosi, F., Sarrazin, F., and Wagener, T., "A Matlab toolbox for Global Sensitivity Analysis," Environmental Modelling & Software, vol. 70, pp. 80–85, Aug. 2015.

doi: 10.1016/j.envsoft.2015.04.009

[69] Janczak, A., *Identification of Nonlinear Systems Using Neural Networks and Polynomial Systems*, Springer, Berlin Heidelberg, Germany, 2005.